\documentclass[12pt,preprint]{aastex}

\slugcomment{Accepted to ApJ}

\shorttitle{CO 3-2 in Virgo galaxies}
\shortauthors{Wilson et al.}

\begin{document}


\title{The JCMT Nearby Galaxies Legacy Survey I. Star Forming Molecular Gas
  in Virgo Cluster Spiral Galaxies}


\author{C. D. Wilson\altaffilmark{1}, B. E. Warren\altaffilmark{1},
F. P. Israel\altaffilmark{2}, 
S. Serjeant\altaffilmark{3},
G. Bendo\altaffilmark{4},
E. Brinks\altaffilmark{5},
D. Clements\altaffilmark{4}, 
S. Courteau\altaffilmark{6}, 
J. Irwin\altaffilmark{6},
J. H. Knapen\altaffilmark{7}, 
J. Leech\altaffilmark{8}, 
H. E. Matthews\altaffilmark{9}, 
S. M\"uhle\altaffilmark{10}, 
A. M. J. Mortier\altaffilmark{11}, 
G. Petitpas\altaffilmark{12}, 
E. Sinukoff\altaffilmark{1}, 
K. Spekkens\altaffilmark{13}, 
B. K. Tan\altaffilmark{8}, 
R. P. J. Tilanus\altaffilmark{14,15},
A. Usero\altaffilmark{5,16}, 
P. van der Werf\altaffilmark{2}, 
T. Wiegert\altaffilmark{17}, 
\& 
M. Zhu\altaffilmark{9,14}
} 

\altaffiltext{1}{Department of Physics \& Astronomy, McMaster  University,
Hamilton, Ontario L8S 4M1 Canada; wilson@physics.mcmaster.ca,
bwarren@physics.mcmaster.ca, sinukoej@mcmaster.ca}
\altaffiltext{2}{Sterrewacht Leiden, Leiden University, PO Box 9513,
  2300 RA Leiden, The Netherlands
  ;israel@strw.leidenuniv.nl, pvdwerf@strw.leidenuniv.nl} 
\altaffiltext{3}{Department of Physics \& Astronomy, The Open
  University, Milton Keynes, Mk7 6AA England }
\altaffiltext{4}{Astrophysics Group, Imperial College London,
Blackett Laboratory, Prince Consort Road, London SW7 2AZ United
Kingdom;
g.bendo@imperial.ac.uk, d.clements@imperial.ac.uk}
\altaffiltext{5}{Centre for Astrophysics Research, University of Hertfordshire,
College Lane, Hatfield  AL10 9AB, United Kingdom;
E.Brinks@herts.ac.uk} 
\altaffiltext{6}{Department of Physics, Engineering Physics and Astronomy, Queen's University, Kingston, ON, Canada;
  irwin@astro.queensu.ca, courteau@astro.queensu.ca} 
\altaffiltext{7}{Instituto de Astrof\'isica de Canarias,
 38200 La Laguna, Spain; jhk@iac.es}
\altaffiltext{8}{Department of Physics, University of Oxford, Keble
  Road, Oxford OX1 3RH, UK; jxl@astro.ox.ac.uk}
\altaffiltext{9}{National Research Council Canada, Herzberg Institute
  of Astrophysics, DRAO, P.O. Box 248, White Lake Road, Penticton, BC
  V2A 6J9, Canada; 
  henry.matthews@nrc-cnrc.gc.ca} 
\altaffiltext{10}{Joint Institute for VLBI in Europe,
Postbus 2, 7990 AA Dwingeloo, The Netherlands; muehle@jive.nl}
\altaffiltext{11}{SUPA\footnote{Scottish Universities Physics
    Alliance},  Institute for
Astronomy, University of Edinburgh, Royal Observatory, Blackford Hill,
Edinburgh, EH9 3HJ, UK; ajm@roe.ac.uk}
\altaffiltext{12}{Harvard-Smithsonian Center for Astrophysics,
  Cambridge, MA 02138; gpetitpa@cfa.harvard.edu}
\altaffiltext{13}{Department of Physics, Royal Military College of
  Canada, PO Box 17000, Station ``Forces'', Kingston, K7K 4B4
  Ontario, Canada; Kristine.Spekkens@rmc.ca}
\altaffiltext{14}{Joint Astronomy Centre, 660 N. A'ohoku Pl., Hilo,
  Hawaii, 96720, USA; 
  r.tilanus@jach.hawaii.edu, m.zhu@jach.hawaii.edu} 
\altaffiltext{15}{Netherlands Organisation for Scientific Research, The Hague}
\altaffiltext{16}{Observatorio Astron\'omico Nacional, C/ Alfonso
 XII 3, 28014 Madrid, Spain;a.usero@oan.es}
\altaffiltext{17}{Department of Physics and Astronomy, University of
  Manitoba, Winnipeg, Manitoba R3T 2N2, Canada;
  wiegert@physics.umanitoba.ca} 
\begin{abstract}
We present large-area maps of the CO $J$=3-2 emission obtained at the
James Clerk Maxwell Telescope for four spiral galaxies in the Virgo
Cluster. We combine these
data with published CO $J$=1-0, 24 $\mu$m, and H$\alpha$ images to
measure the CO line ratios, molecular gas masses, and instantaneous
gas depletion times. For three galaxies in our sample (NGC 4254,
NGC4321, and NGC 4569), we obtain molecular gas masses of $7\times
10^8 - 3\times 10^9$
M$_\odot$ and disk-averaged instantaneous gas depletion times of 
1.1-1.7 Gyr. We argue that the CO $J$=3-2 line is a better tracer of
the dense star
forming molecular gas than the CO $J$=1-0 line, as it shows a better
correlation with the star formation rate surface density
both within and between galaxies. NGC 4254 appears to have
a  larger star formation efficiency(smaller gas depletion time),
perhaps because it is on its first passage through the Virgo
Cluster. NGC 4569 shows a large-scale gradient in the gas properties
traced by the CO $J$=3-2/$J$=1-0 line ratio, which suggests that its
interaction with the intracluster medium is affecting the dense
star-forming portion of the interstellar medium directly. The fourth
galaxy in our sample, NGC 4579, has weak CO $J$=3-2
emission despite having bright 24 $\mu$m emission; however, much of
the central luminosity in this galaxy may be due
to the presence of a central AGN.
\end{abstract}



\section{Introduction}

Molecular gas provides the fuel for star formation and therefore
plays an important role in both the current appearance and future
evolution of galaxies. Early surveys for molecular gas in 
galaxies revealed a wide range in total gas content, from
the rarely detected elliptical galaxies to gas-rich massive spirals
such as M51 \citep{ys91,b93,y95}. For spiral galaxies, the variations in gas
content roughly track the star formation 
rate for different systems, such that the global star formation efficiency
(the inverse of which is the gas depletion
time, $t_{gas} = M_{H_2} / SFR$) does not depend strongly on morphology 
when far-infrared luminosity is used to trace the
star formation rate \citep{ys91}. Detailed studies of individual galaxies
provided evidence for 
enhanced star formation efficiencies 
(implying shorter instantaneous gas depletion times)
at certain positions within
spiral arms \citep{v88} 
or along them \citep{k92,k96} compared to the disk-averaged value.
A focused survey of spiral galaxies in the Virgo cluster showed that
the molecular gas content of HI-deficient cluster spirals
is similar to that of their more gas-rich counterparts 
\citep{ky89}.
\citet{kenn07} find that the star formation rate surface density
correlates with the molecular but not the atomic gas surface density.

To date, most large surveys of molecular gas in galaxies have been
carried out using the CO $J$=1-0 transition; this line is sensitive to
the vast majority of molecular gas in the cold interstellar medium in
galaxies, since the $J$=1 level is only 5.5 K above the ground state and
the $J$=1-0 transition has a critical density of $1.1\times 10^3$ cm$^{-3}$ in optically thin gas.
The detailed connection between CO emission and molecular gas mass and
ultimately the star formation process depends rather critically on the
value of the CO-to-H$_2$ conversion factor \citep{s88} and any
dependencies of that conversion factor on temperature, density, or
metallicity \citep{dss85,mb88}.
While individual galaxies have been studied in higher transitions,
primarily CO $J$=2-1 and CO $J$=3-2, the only large surveys
available to date in the CO $J$=3-2 transition are observations of
the central regions of galaxies \citep{m99,hs03,y03,komugi07}.
Since the CO $J$=3 level is  33 K above the ground
state and the $J$=3-2 transition 
has a critical density of $2.1\times 10^4$ cm$^{-3}$ in optically thin
gas, the $J$=3-2
transition naturally traces 
gas that is on average warmer and/or denser than the CO $J$=1-0 line. 

Examining the CO $J$=3-2/1-0 line ratio both within and between galaxies
can give 
information on any large-scale variations in the physical conditions in
the molecular gas \citep{w97,pw00}.
In addition, the CO $J$=3-2 emission may trace the molecular gas that is
more directly connected with star formation; indeed, \citet{i08} have
found that the global CO $J$=3-2 luminosity correlates nearly linearly with the
global star formation rate over five orders of magnitude. 
\citet{b08} also find a nearly linear relation between the H$_2$ surface
density traced by the CO $J$=2-1 line and the star formation rate surface
density. These relations are similar to the linear correlation of HCN
with far-infrared luminosity seen by \citet{gao04} and suggests that
these CO transitions may also be effective tracers of the densest, star
forming gas. In contrast,
the CO $J$=1-0 luminosity  has a much shallower correlation with
star formation rate  with a slope $0.62 \pm 0.08$ \citep{g05}.
This shallower slope implies that there is not a one-to-one relation
between the amount of molecular gas (as traced by the CO $J$=1-0 line) and
the amount of star formation in a galaxy. The shallower slope compared
to the CO J=3-2 data also indicates that the two lines are not tracing
exactly the same gas components.
For example, the CO $J$=1-0 line can include
emission from lower opacity, more diffuse molecular gas
\citep{ww94,rb05}.
Examining the correlation between molecular gas and star formation
rate as a function of the gas density may yield new insights into the
star formation process in galaxies.

An ideal laboratory for the acquisition of resolved molecular maps
in galaxies is the nearby Virgo Cluster.  At a distance of 16.7 Mpc
\citep{mei07}, it is the largest concentration of galaxies
\citep[cluster mass $4.2 \times 10^{14}$ M$_\odot$;][]{mcl99}
within 35 Mpc. It 
has been the target of numerous studies over the years and plays an
important role in our understanding of how galaxies evolve in dense
environments. The most extensive optical survey was carried out by
\citet{b85} and resulted in the Virgo Cluster Catalog (VCC) containing
$\sim 2100$ certain or possible cluster members over an area of 140
deg$^2$. The Virgo Cluster is particularly interesting for studying
the impact of the dense environment on the ISM in galaxies, as
properties such as ram-pressure stripping, tidal interactions,
etc., become important 
in environments with a massive intracluster medium.

Recently, the Virgo Cluster has been included
within the area being mapped in the HI 21 cm line by the ALFALFA
survey \citep{gio05}. 
Comparison with
optical catalogs shows $\sim 200$ detections associated with optical
counterparts 
as well as 10 detections associated with tidal debris
tails \citep{gav08}. High-resolution HI observations with the VLA have
identified 
seven spiral galaxies with long HI tails pointing away from M87 \citep{chu07},
which suggests that interaction with the cluster as a whole has
created these tails. Other evidence for environmental modification
includes ram-pressure stripped extraplanar gas tails and disturbances
in radio morphologies \citep{k08,m08}.

Spiral galaxies in the Virgo Cluster were first surveyed for molecular
emission by \citet{s86}, who detected 23 out of 47 spirals observed
with a beam size of 100$^{\prime\prime}$. 
The primary 
reference for the molecular gas content of spiral galaxies in the
Virgo Cluster remains the work of \citet{ky86,ky88a,ky88b,ky89}.
Using a complete blue-magnitude selected sample, 
they found that the molecular gas content of HI-deficient cluster spirals
in Virgo is similar that of to their more gas-rich counterparts.
Interferometric CO $J$=1-0 observations of 15 Virgo galaxies
show a wide range of molecular gas morphology
as well as higher molecular gas concentrations in the galaxies with
deeper gravitational potential wells \citep{s03a,s03b}.
High resolution CO $J$=1-0 images for a few galaxies in the Virgo
cluster are also presented in \citet{h03} and \citet{k07}.
\citet{hs03} made low resolution observations of the CO $J$=3-2 and
$J$=2-1 lines with the KOSMA 3~m telescope of the 20 strongest sources
from the CO $J$=1-0 survey of \citet{s86}. They detected
18 and 16 spiral galaxies, respectively, in the two lines and measured
CO $J$=3-2/$J$=1-0 line ratios ranging from 0.14 to 0.35  and CO $J$=2-1/1-0
line ratios from 0.5 to 1.1.



The star formation properties of Virgo cluster galaxies have also been
 studied extensively.
\citet{ky04a} found that half of the bright Virgo spirals have truncated
H$\alpha$ disks even though they have relatively undisturbed stellar
disks. This truncation suggests that ICM-ISM stripping is primarily
responsible 
for the reduced star formation rates of bright Virgo spirals
\citep{g02,ky04b}. \citet{fg08} found that cluster members which have
lost even a moderate amount of their atomic gas also have lower molecular
gas content and star formation rates.

In this paper, we present some first results from the JCMT Nearby
Galaxies Legacy Survey (NGLS).
We focus on new, wide-area observations of the CO $J$=3-2
emission for the four large
spiral galaxies in the Virgo Cluster 
(\object{NGC 4254}, \object{NGC 4321}, \object{NGC 4569}, and
\object{NGC 4579}) that are part of the Spitzer Infrared Nearby
Galaxies Survey (SINGS) sample \citep{k03}. We briefly
describe the structure and goals of the NGLS in \S\ref{sec-ngls} and
review previous observations of the interstellar medium 
in the particular Virgo Cluster galaxies in our sample in \S\ref{sec-bkgd}.
We discuss our
observations and data reduction in 
\S\ref{sec-obs}. 
In \S\ref{sec-lineratios}, we combine the data with CO $J$=1-0 maps from
\citet{k07} to estimate the CO $J$=3-2/$J$=1-0 line ratio and the
molecular gas mass.
In \S\ref{sec-tgas},
we combine these new data with H$\alpha$
data and reprocessed 24 $\mu$m data from the SINGS survey \citep{k03} 
to obtain maps of the star formation rate \citep{c07} and gas
depletion time.
We discuss the usefulness of the CO $J$=3-2 line as a tracer for dense star
forming gas and the gas and star formation properties of the four
galaxies in the context of their internal structure and broader
environment in \S\ref{sec-discuss}.
We give our conclusions 
in \S\ref{sec-concl}.

\subsection{The JCMT Nearby Galaxies Legacy Survey}\label{sec-ngls}

The JCMT Nearby Galaxies Legacy Survey
(NGLS)\footnote{http://www.jach.hawaii.edu/JCMT/surveys/} focuses on
the molecular gas 
and dust content of galaxies within 25 Mpc using three primary
samples: (1) galaxies from the SINGS \citep{k03} sample 
\citep[see][for first results from the NGLS for additional SINGS
galaxies]{w08};
(2) an HI flux limited sample of galaxies from the Virgo
cluster; and (3) an HI flux limited sample of galaxies in the field.
The first phase of the NGLS consists of  CO $J$=3-2 observations out
to $D_{25}/2$ for 21 galaxies from the SINGS sample as well as the complete
Virgo sample; this phase of the survey is currently complete. The
second phase consists of CO $J$=3-2 observations for the entire sample
and will begin in 2009 if additional observing time is allocated to
the survey. The third phase, for which observing time has been
allocated already, will consist of SCUBA-2 observations of the
entire sample and should begin in late 2009.

Specific science goals for the NGLS include: searching for evidence of
cold dust and measuring its mass fraction in galaxies of different
types; measuring the amount of warm, dense molecular gas associated
with star formation using the CO $J$=3-2 line; comparing galaxies with
similar morphologies and luminosities in the field and in the Virgo
cluster to determine the effects of cluster membership; using CO
rotation curves to trace the dark matter distribution and the
frequency of occurrence of nuclear gas concentrations which may feed
central starbursts or black holes; and measuring the local
submillimeter luminosity function and dust mass function to
luminosities up to 100 times fainter than previous studies
\citep[e.g.,][]{d00}. 




\subsection{Properties of four Virgo spiral galaxies}\label{sec-bkgd}

Four large
spiral galaxies
(\object{NGC 4254}, \object{NGC 4321}, \object{NGC 4569}, and
\object{NGC 4579}) are the only Virgo Cluster galaxies
included in the SINGS survey 
\citep{k03} and there is a large collection of information on their
dust and interstellar medium properties. Some basic properties of the
galaxies are given in Table~\ref{tbl-props}.
The global 1-850 $\mu$m spectral energy
distributions (SEDs) for these four galaxies 
are very similar \citep{d05}, although the $\alpha$
parameter in the SED fit ranges from a high 3.8 in NGC 4579 to a low of 2.4
in NGC 4254. \citet{d07} calculate the dust mass and gas to dust
ratio for 65 galaxies in the SINGS sample. Of the four Virgo spirals
discussed in our paper,
only NGC 4579 stands out as having a factor of 3 larger gas to dust
ratio; this is the second highest gas to dust ratio in the entire sample
analyzed by \citet{d07}. 
This high gas-to-dust ratio may be a
  by-product of the difficulty in determining
the 850 $\mu$m flux from
  the relatively low signal-to-noise SCUBA map (Bendo, private
  communication).  
Morphologically, of the four Virgo spirals, NGC 4569 is the most 
centrally concentrated at 24 $\mu$m, while NGC 4254 is both the
least centrally concentrated and the most asymmetric \citep{b07}.
The star formation distribution based on
H$\alpha$ emission is classified as normal for NGC 4254 but
truncated/normal for the other three galaxies \citep{ky04a}. Below we
give some important details on each of the individual galaxies.

Recent HI observations of NGC4254 (M99) show a long tidal tail extending to
the north; its lopsided morphology, high HI content, and
high relative velocity suggest that this galaxy is undergoing its
first encounter with the center of the Virgo cluster \citep{h07}. The
galaxy has a normal 
molecular gas to total gas fraction \citep{n06}.

NGC 4321 (M100) is classified as HII/LINER \citep{h97}. High
resolution CO $J$=1-0 observations show two spiral 
arms and gas in the nucleus as well as faint emission bridging the
arms and the nucleus \citep{gb05}. The HI emission is truncated near
the edge of the optical disk \citep{c90,braun07}.
NGC 4321 is interacting with NGC 4322, as shown by the bridge between
them that is visible in optical light but not in HI \citep{k93}. It
also has a rather 
prominent circumnuclear 
ring with much enhanced star formation \citep{k95a,k95b}.

NGC 4569 (M90) has been called an anemic spiral because of the small extent
of its HI emission relative to its optical size
\citep{s99}, although \citet{ky04a} prefer a designation of
truncated/normal because its star formation rate per unit area is
relatively normal. \citet{b07} note that there is almost no 24 $\mu$m
emission from the extended disk.
The CO
$J$=1-0 emission shows a compact bar oriented roughly north-south
\citep{s99} with
complex kinematics indicating non-circular motions in the central
arcminute \citep{j05}. 
This galaxy has a higher 
molecular gas to total gas fraction than NGC 4254, perhaps due to ram
pressure stripping of HI even within the stellar disk or a higher external
pressure due to the intracluster medium \citep{n06}. However, the fact
that the only remaining gas is in the inner disk, which tends to have
a high molecular gas fraction, could also explain this result.
\citet{ky04a} find that the galaxy has some HII regions that appear to
be outside the plane. They also note that rotation plus forward
motion relative to the ICM can create an asymmetric ram pressure,
which in simulations can produce a dominant extraplanar gas arm
emerging from a truncated gas disk. \citet{v04} present also evidence
for ram pressure stripping by comparing the kinematically distinct
western arm with dynamical simulations. \citet{b06} argue that active
gas stripping, rather than simply stopping gas inflow, is required to
explain the star formation history in NGC 4569. \citet{j05}
suggest that NGC 4569 is in the early stages of bar or tidally driven
gas inflow.

NGC 4579 (M58) is classified as LINER 1.9/Seyfert 1.9
\citep{gb05}. High resolution CO $J$=2-1 observations show two spiral 
arms approaching to within a few arcseconds of the AGN \citep{gb05}
and weak
CO emission at the AGN itself \citep{gb08}. At lower
resolution, weak CO emission is detected throughout the disk
\citep{n06}. Despite
the weak emission, this galaxy has a higher 
molecular gas to total gas fraction than NGC 4254, and so may
also have been affected by ram
pressure stripping or a higher external
pressure \citep{n06}.

In summary, this small sample of four spiral galaxies probes a
large fraction of the characteristics seen for luminous spirals in the
Virgo cluster such as HI deficiency, star formation rate, and presence
of nuclear 
activity. Thus, a detailed study of the 
molecular gas and star formation
properties of these galaxies should provide 
insight into the effects on
the ISM in galaxies
and galaxy structure and evolution that can be addressed using the
complete Virgo sample from the NGLS. 

\section{Observations and Data Processing}\label{sec-obs}

\subsection{CO $J$=3-2 Data}

The CO $J$=3-2 observations were obtained as part of the JCMT Nearby
Galaxies Legacy Survey
(NGLS). We used the
16 pixel array receiver 
HARP-B  with the ACSIS correlator  configured to have a
bandwidth of 1 GHz and a resolution of 0.488 MHz (0.43 km s$^{-1}$
at the frequency of the CO $J$=3-2 transition). The galaxies were
observed in raster mapping mode to cover a rectangular area
corresponding to $D_{25}/2$ with a 1 sigma sensitivity of better than
19 mK ($\rm T_A^*$) at a spectral resolution of 20 km
s$^{-1}$.  The angular resolution of the JCMT at this frequency is
14.5$^{\prime\prime}$ and the images were processed using
7.2761$^{\prime\prime}$ pixels (the recommended pixel size in the
reconstruction of HARP raster maps).
Details of the observations are given in Table~\ref{tbl-obs}.

\begin{figure}
\includegraphics[angle=0,scale=.4]{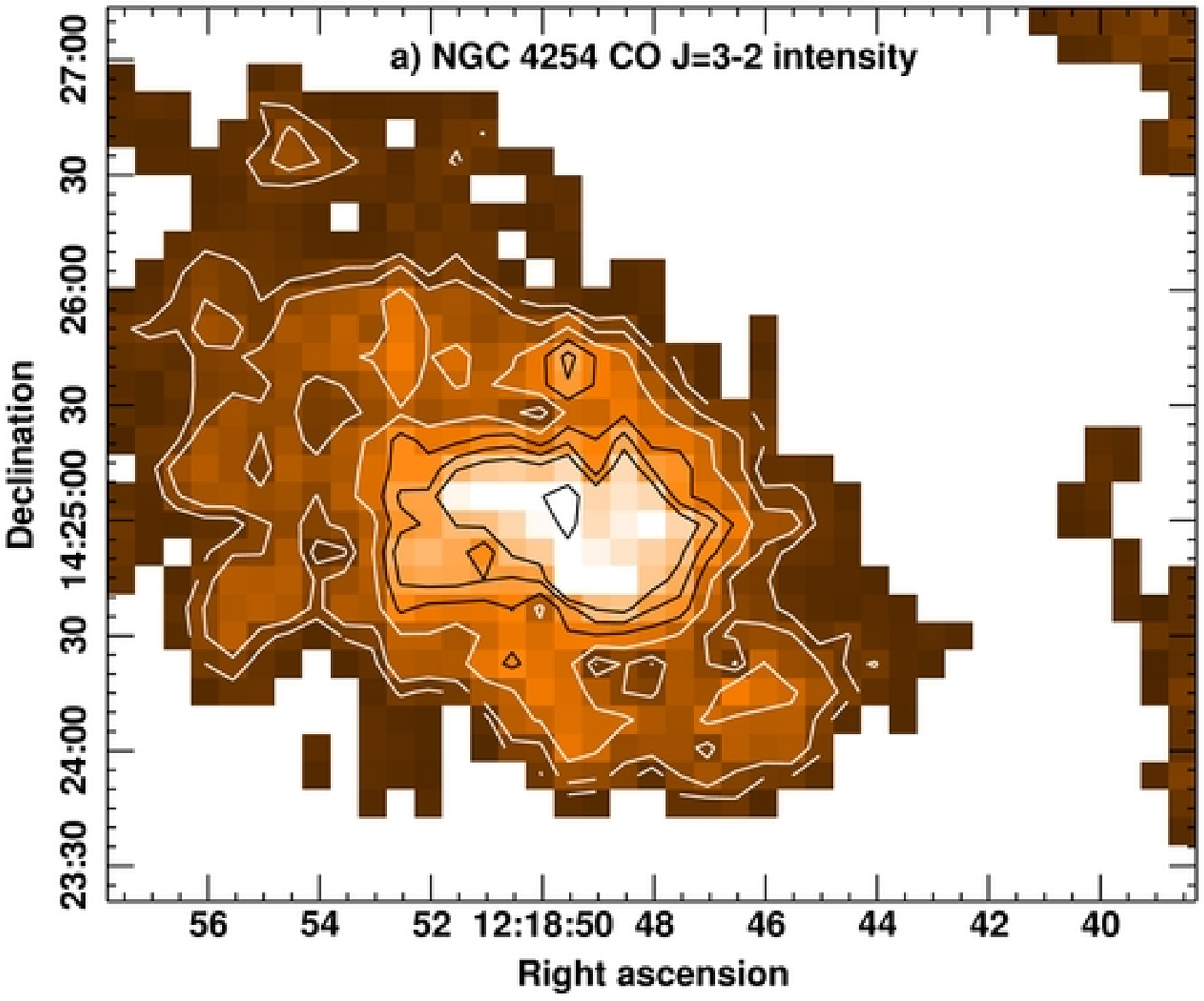}
\includegraphics[angle=0,scale=.4]{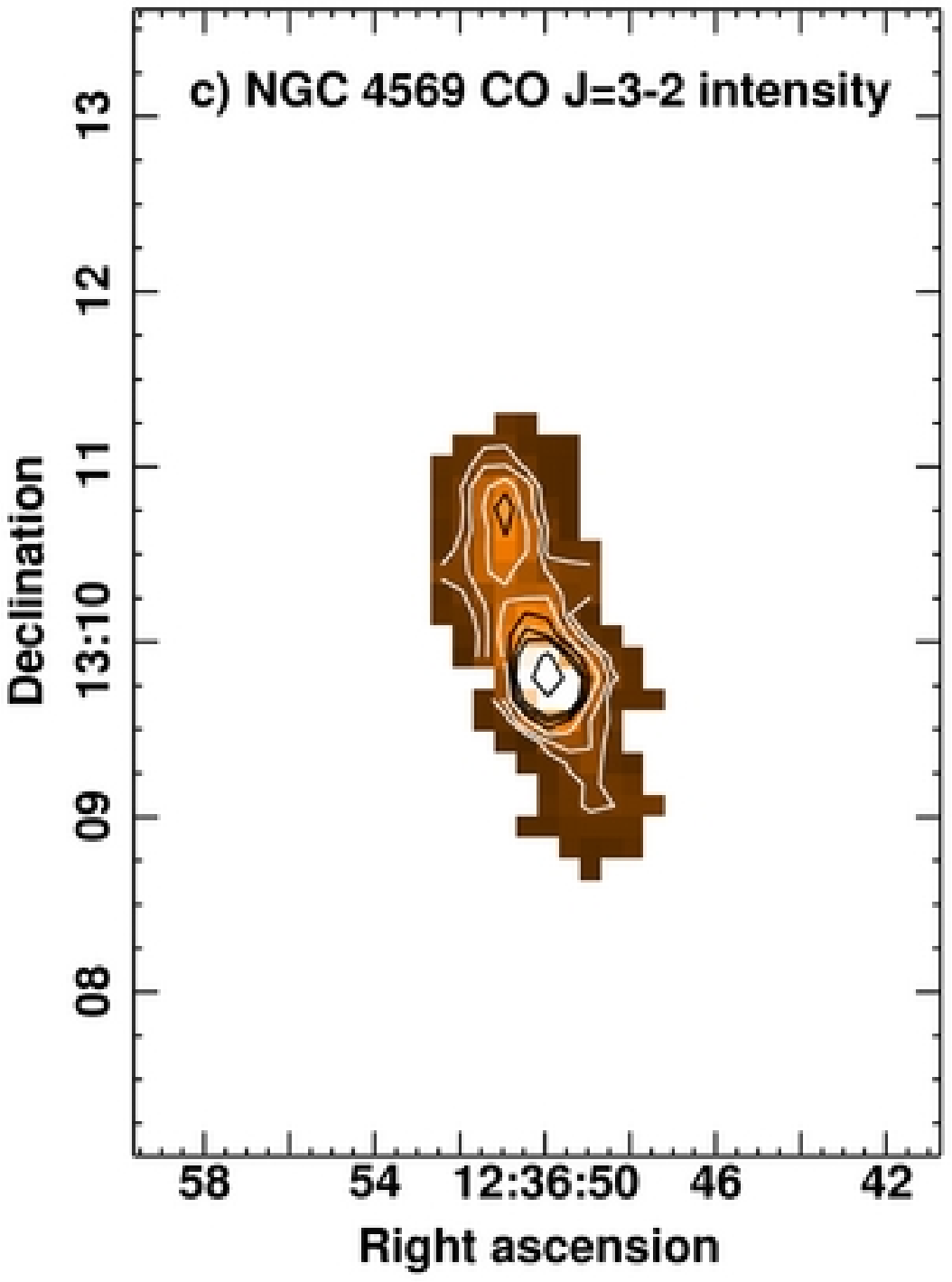}
\includegraphics[angle=0,scale=.4]{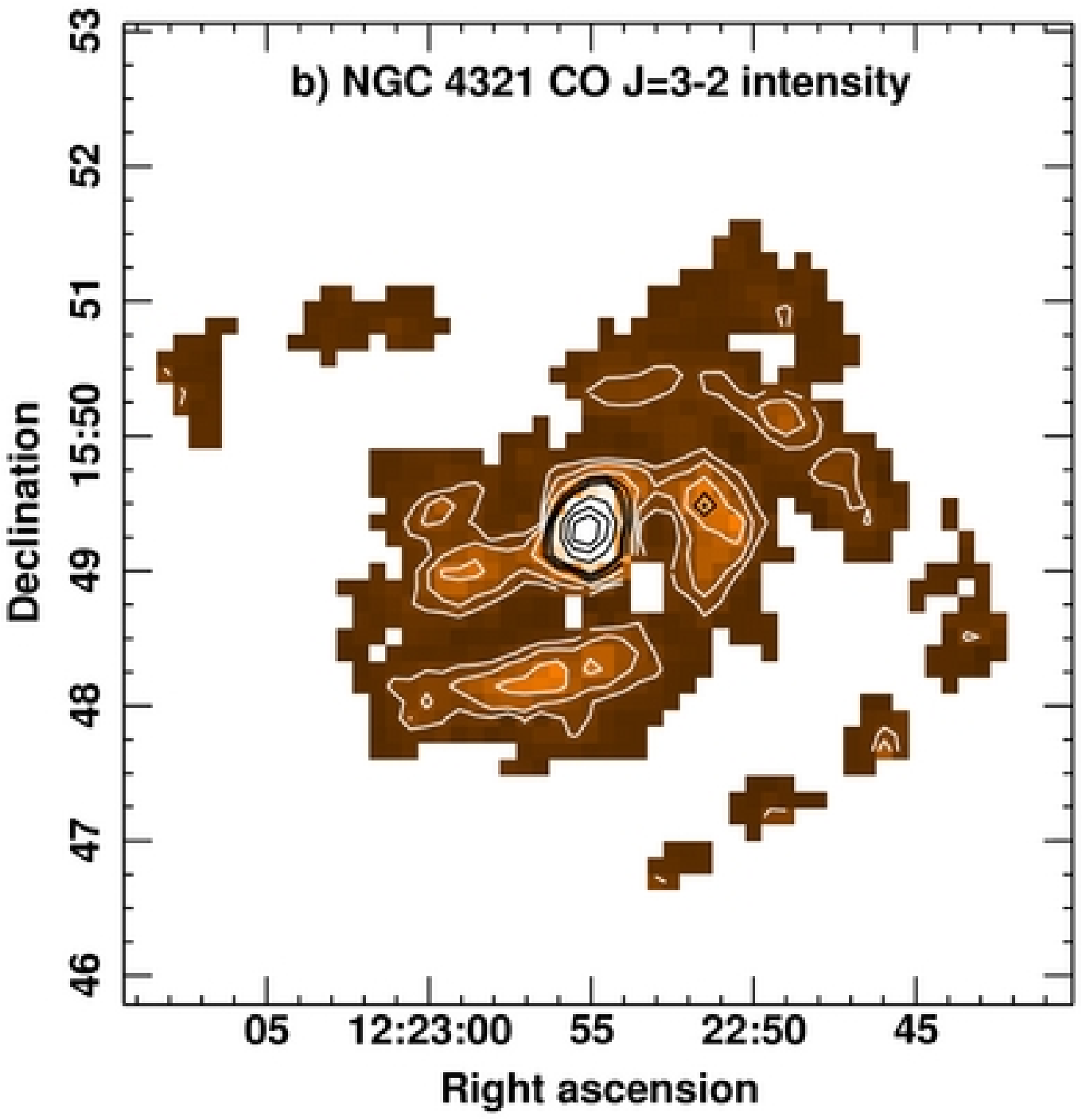}
\includegraphics[angle=0,scale=.4]{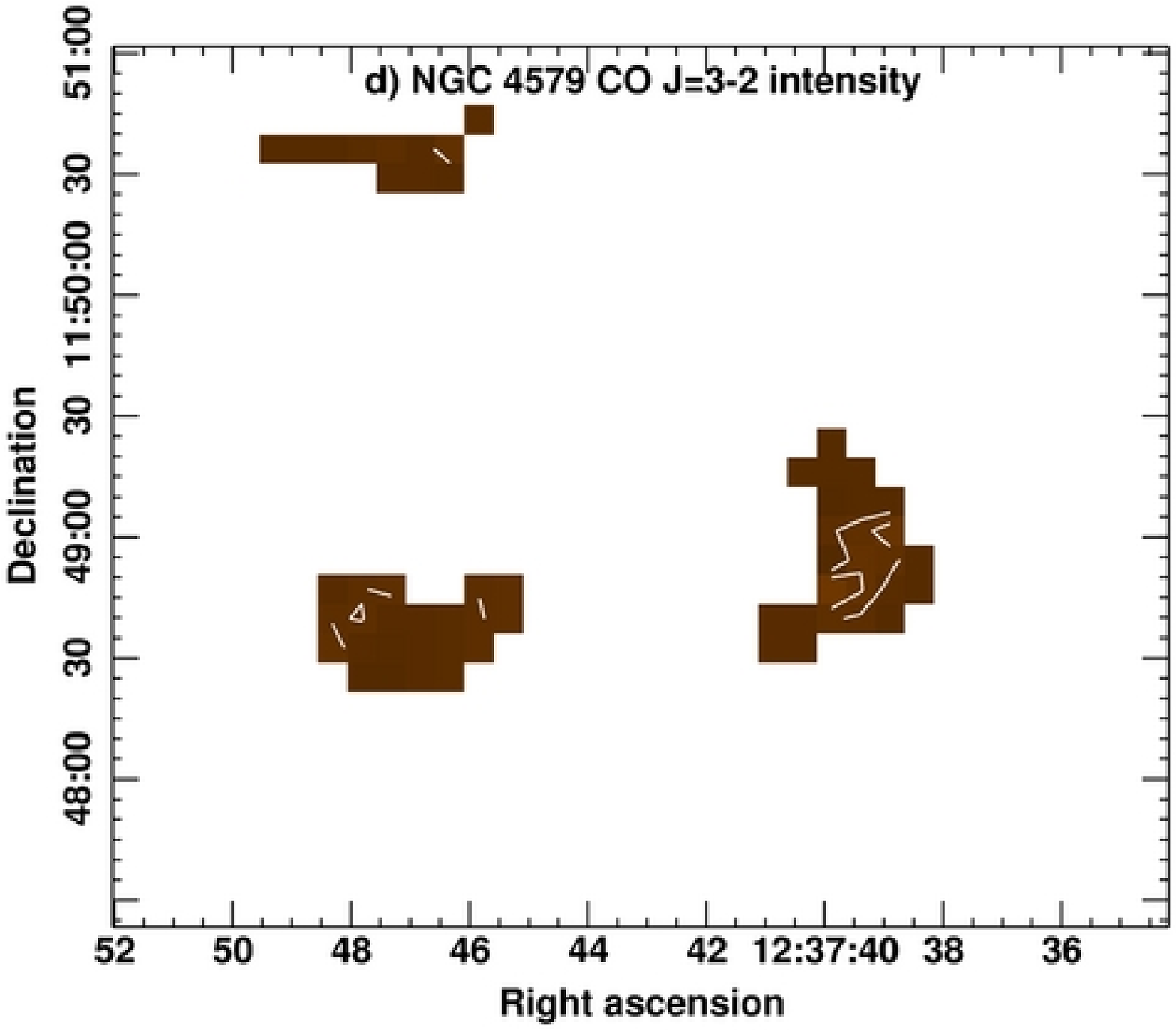}
\caption{(a) CO $J$=3-2 integrated intensity for NGC 4254.
Contours
  are 1,2,4,6,8,10,20 K km s$^{-1}$ and greyscale runs from -3 to 15 K
  km s$^{-1}$. Coordinate epoch is J2000.
(b)  CO $J$=3-2 integrated intensity for NGC 4321. Contours
  are 1,2,4,6,8,10,20,30,40,50
K km s$^{-1}$ and greyscale runs from -3 to 15 K km s$^{-1}$. 
(c)  CO $J$=3-2 integrated intensity  for NGC 4569. 
Contours
  are 1,2,4,6,8,10,20 K km s$^{-1}$ and greyscale runs from -3 to 15 K km s$^{-1}$.
(d)  CO $J$=3-2 integrated intensity for NGC 4579. 
Contours
  are 0.25,0.5 K km s$^{-1}$ and greyscale runs from -3 to 15 K km s$^{-1}$.
\label{fig-co32}}
\end{figure}

\begin{figure}
\includegraphics[angle=0,scale=.4]{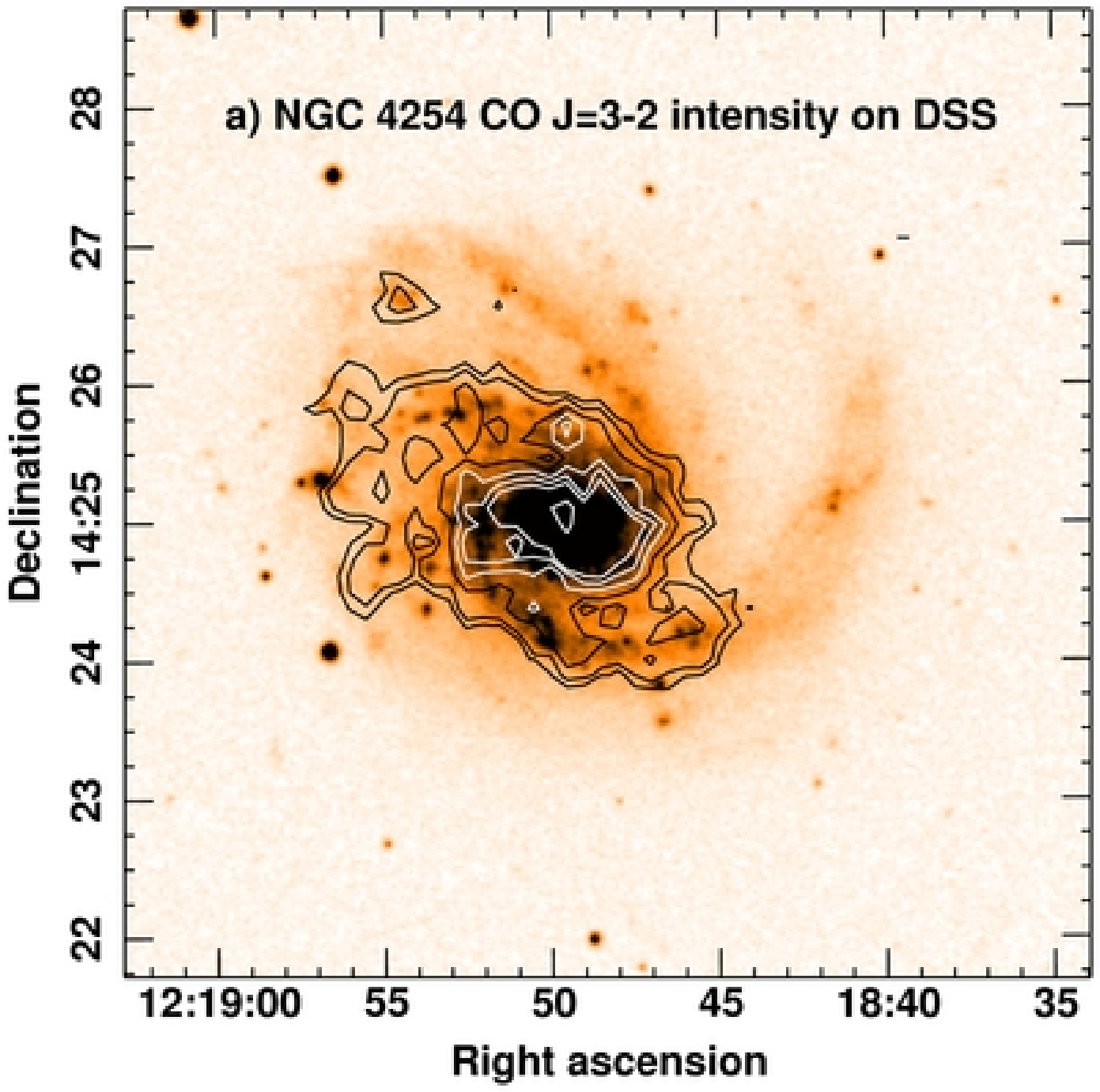}
\includegraphics[angle=0,scale=.4]{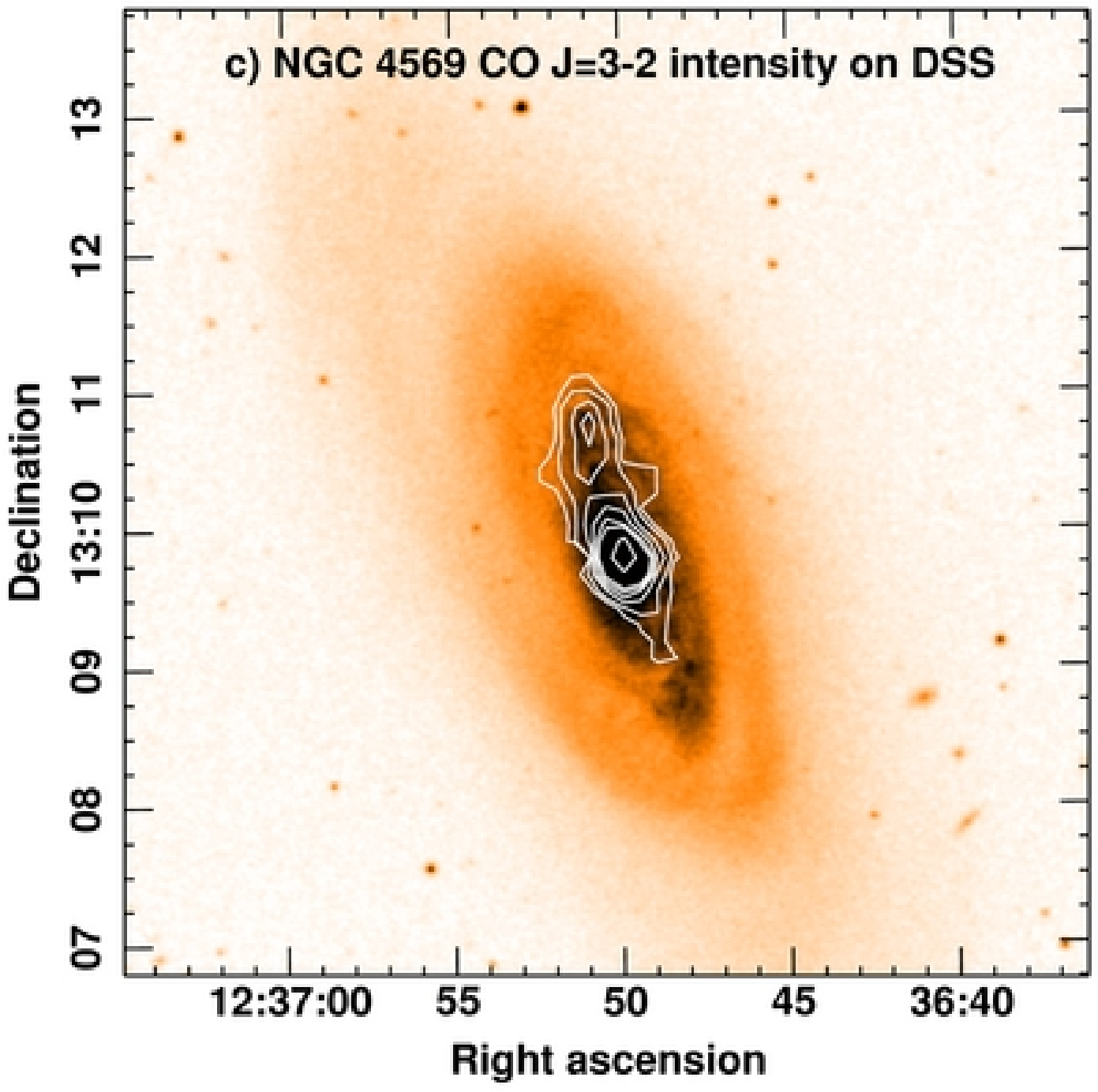}
\includegraphics[angle=0,scale=.4]{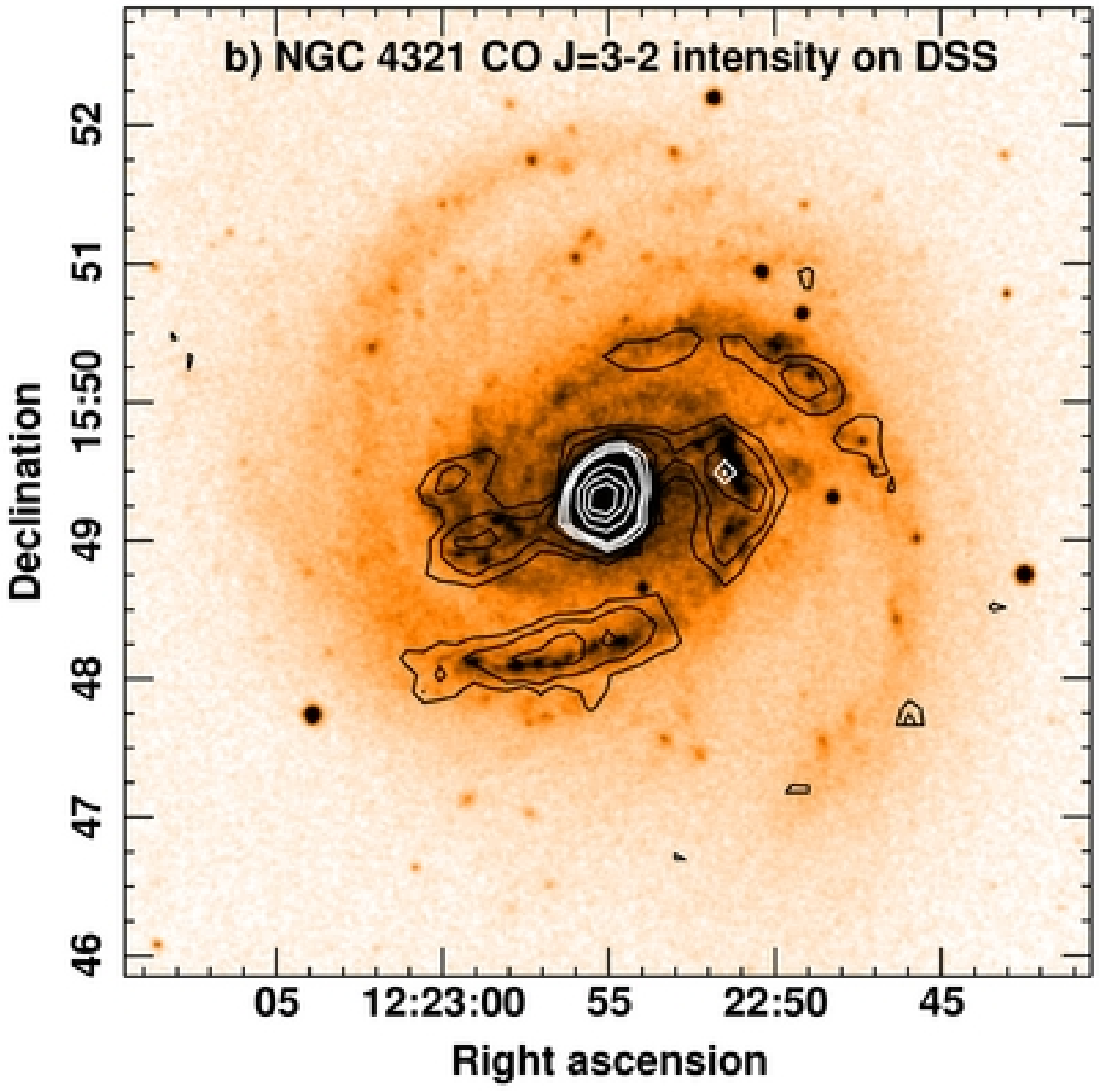}
\includegraphics[angle=0,scale=.4]{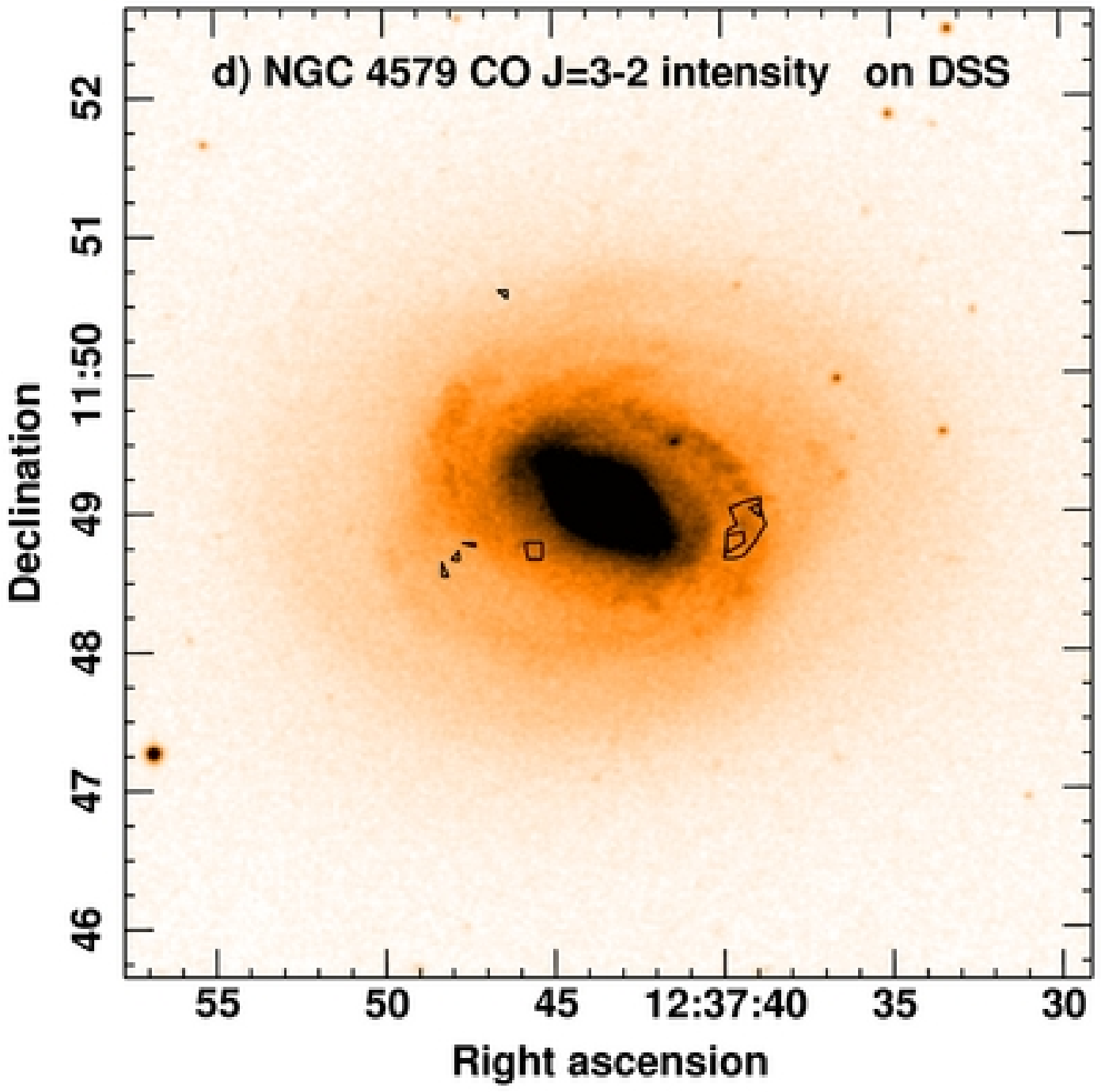}
\caption{(a) CO $J$=3-2 integrated intensity for NGC 4254 overlaid on an
  image from the Digitized Sky Survey. 
Contours
  are 1,2,4,6,8,10,20 K km s$^{-1}$.  Coordinate epoch is J2000.
(b) CO $J$=3-2 integrated intensity for NGC 4321 overlaid on an
  image from the Digitized Sky Survey. 
Contours
  are 1,2,4,6,8,10,20,30,40,50
K km s$^{-1}$. 
(c) CO $J$=3-2 integrated intensity for NGC 4569 overlaid on an
  image from the Digitized Sky Survey. 
Contours
  are 1,2,4,6,8,10,20 K km s$^{-1}$. 
(d) CO $J$=3-2 integrated intensity for NGC 4579 overlaid on an
  image from the Digitized Sky Survey. 
Contours
  are 0.25,0.5 K km s$^{-1}$. 
\label{fig-co32_onDSS}}
\end{figure}

Details of the reduction of the CO $J$=3-2 data are given in
\citet{w08} and so we give only a brief summary here. The
individual raw data files were flagged to remove data from
any of the 16 individual receptors  with bad
baselines and then the scans were combined into a data cube using a
${\rm sinc}(\pi x){\rm sinc}(k\pi x)$ kernel as the weighting function
to determine the contribution of individual receptors to each pixel in
the final map. A mask was created to identify
line-free regions of the data cube and a third-order baseline was fit
to those line-free regions. The clumpfind
algorithm \citep{w94} implemented as part of the
Starlink/cupid\footnote{Cupid and kappa are part of the Starlink 
  software package, which is available for download from
http://www.jach.hawaii.edu} task 
findclumps was used to identify regions with emission above three
times the rms noise in a data cube that had been boxcar smoothed
by 3 pixels and 25 velocity channels. Moment maps were created from
the original data cube using the mask created by findclumps.
The data were converted to the main beam temperature scale by dividing
by $\eta_{MB}=0.67$. Maps of the CO $J$=3-2 integrated intensity
are given in Figure~\ref{fig-co32} and are 
overlaid on images from the Digitized Sky Survey in
Figure~\ref{fig-co32_onDSS}.

\subsection{Ancillary data}

The CO $J$=1-0 first moment maps from \citet{k07} were downloaded from
their survey web site. 
These CO $J$=1-0 data are already
in the $T_{mb}$ temperature scale.
The angular resolution of the JCMT in the CO $J$=3-2 line closely
matches that of the Nobeyama 45 m telescope in the CO $J$=1-0 line. Thus,
no additional smoothing was applied.

For comparison with the maps from \citet{k07},
the CO $J$=1-0 first moment maps from \citet{h03} were downloaded from the
NCSA Astronomy Digital Image Library; there was no CO $J$=1-0 image for
NGC 4254. The image for NGC 4579 did not include any single dish
(short-spacing) data \citep{h03}
and so must be considered a lower limit to the
total CO $J$=1-0 intensity. Each image was convolved with
the appropriate two-dimensional gaussian to give a smoothed image with
a 14.5$^{\prime\prime}$ circular gaussian beam to match the JCMT
resolution and converted from Jy beam$^{-1}$ km s$^{-1}$ to K
km s$^{-1}$ using a conversion factor of 0.438 K Jy$^{-1}$.

The 24 $\mu$m data from the
SINGS survey were reprocessed using an updated version of the data
reduction description given in the SINGS Fourth Data Delivery User's
Guide \citep{SINGS}
with three modifications.  
First, a background offset related to the scan mirror position has been
measured and removed from the data.
Second, asteroids have been identified and removed
from the data.  Third, the updated flux calibration terms given by
\citet{e07}
have been applied to the data.  To compare
the 24 $\mu$m images to the JCMT data, we convolved the 24 $\mu$m
images with special convolution kernels that match the profiles of the
PSF to the 14.5$^{\prime\prime}$ gaussian function that describes the
JCMT beam. 
These convolution kernels were created using the emprical 24 $\mu$m
PSF described by \citet{y09} and the
prescription given by \citet{g08}.

The H$\alpha$ images were downloaded from the SINGS
website\footnote{http://sings.stsci.edu} 
\citep{k03}; there was no image for NGC 4569, so we used
the image from the GOLD Mine Database \citep{g03}.
Although the images were background subtracted, a weakly varying
background across the image affected the flux levels once the images
were convolved to match the JCMT resolution. 
We therefore binned the image by 100 pixels and clipped any data above
2 sigma to remove real emission, and then fit a
two-dimensional first-order polynomial to the resulting image.
The data were calibrated as described in the SINGS release notes.
The images were corrected for contamination
by the two [NII] lines using the [NII]/H$\alpha$ ratios from
\citet{p07}. 
Finally, we convolved the images with a 14.5$^{\prime\prime}$ gaussian
to match the resolution of the JCMT.

\begin{figure}
\includegraphics[angle=0,scale=.3]{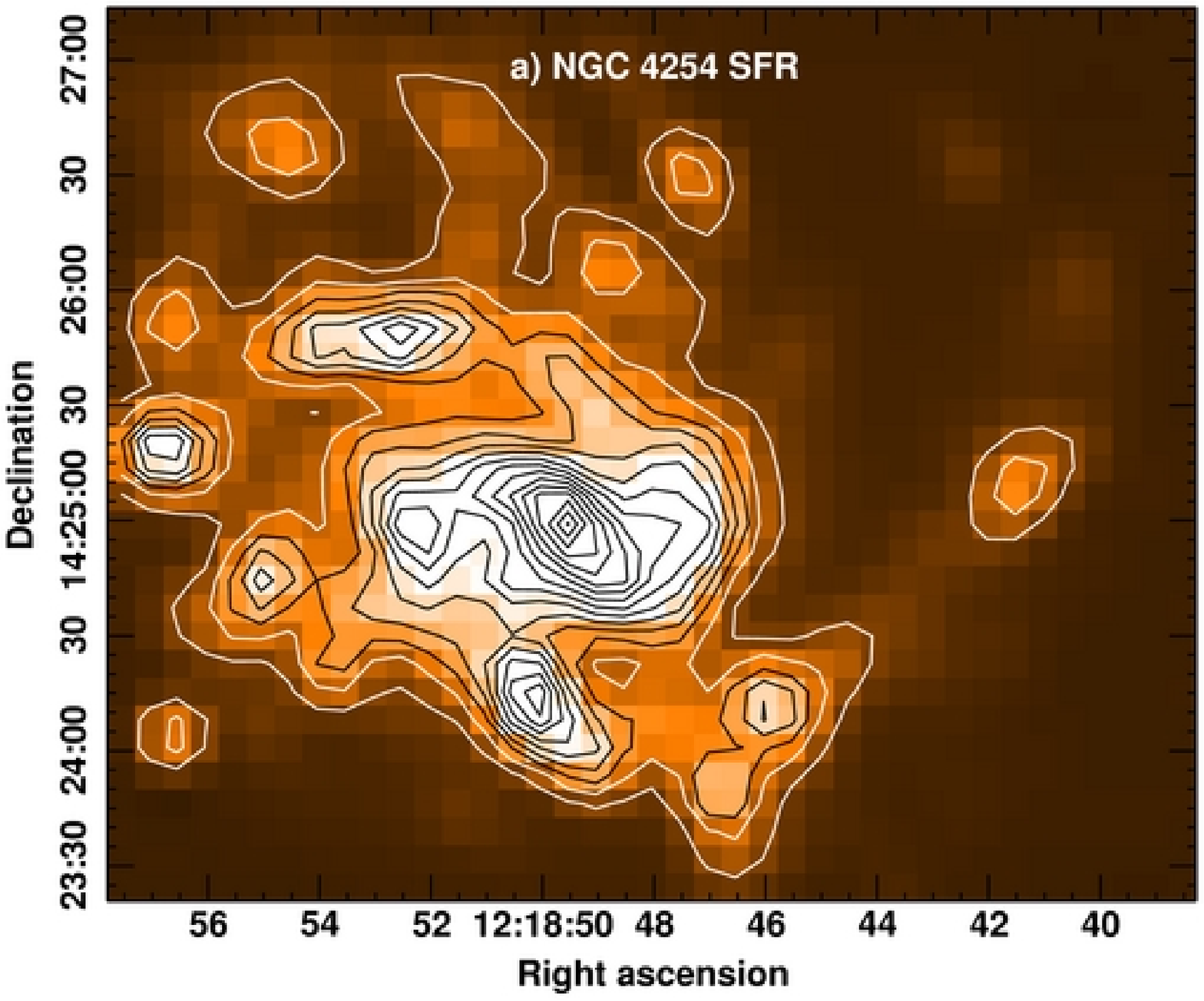}
\includegraphics[angle=0,scale=.3]{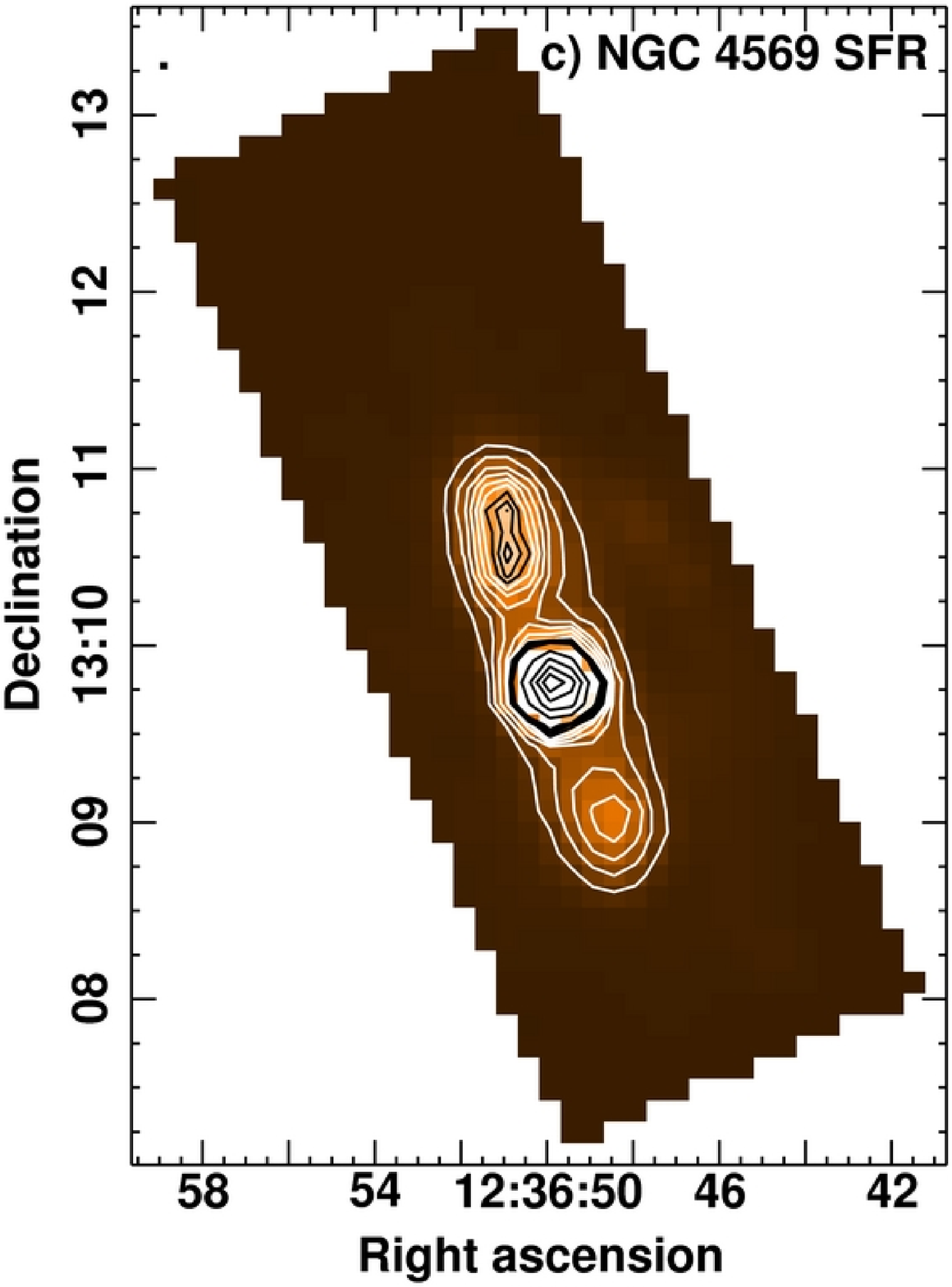}
\includegraphics[angle=0,scale=.2]{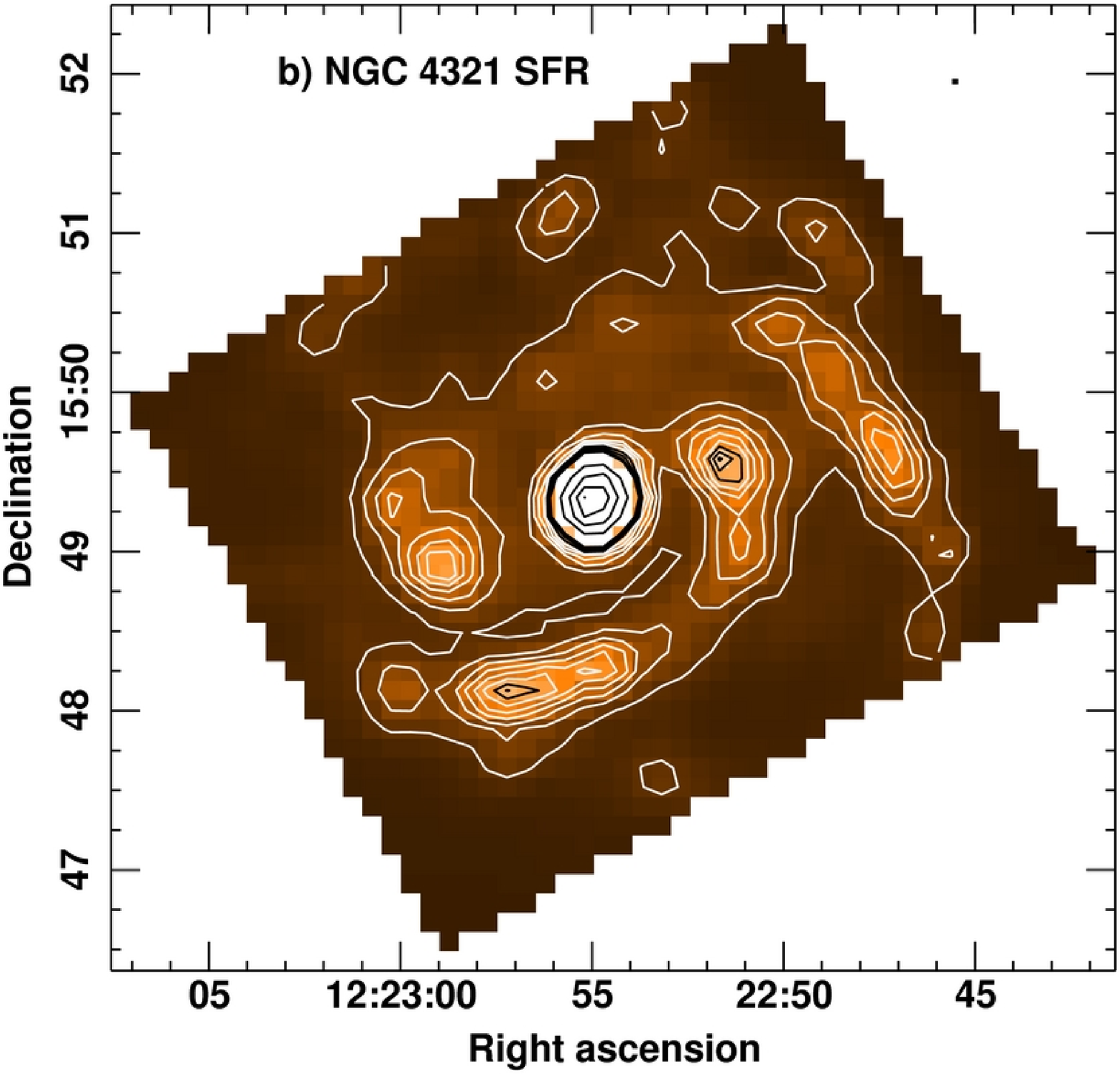}
\includegraphics[angle=0,scale=.2]{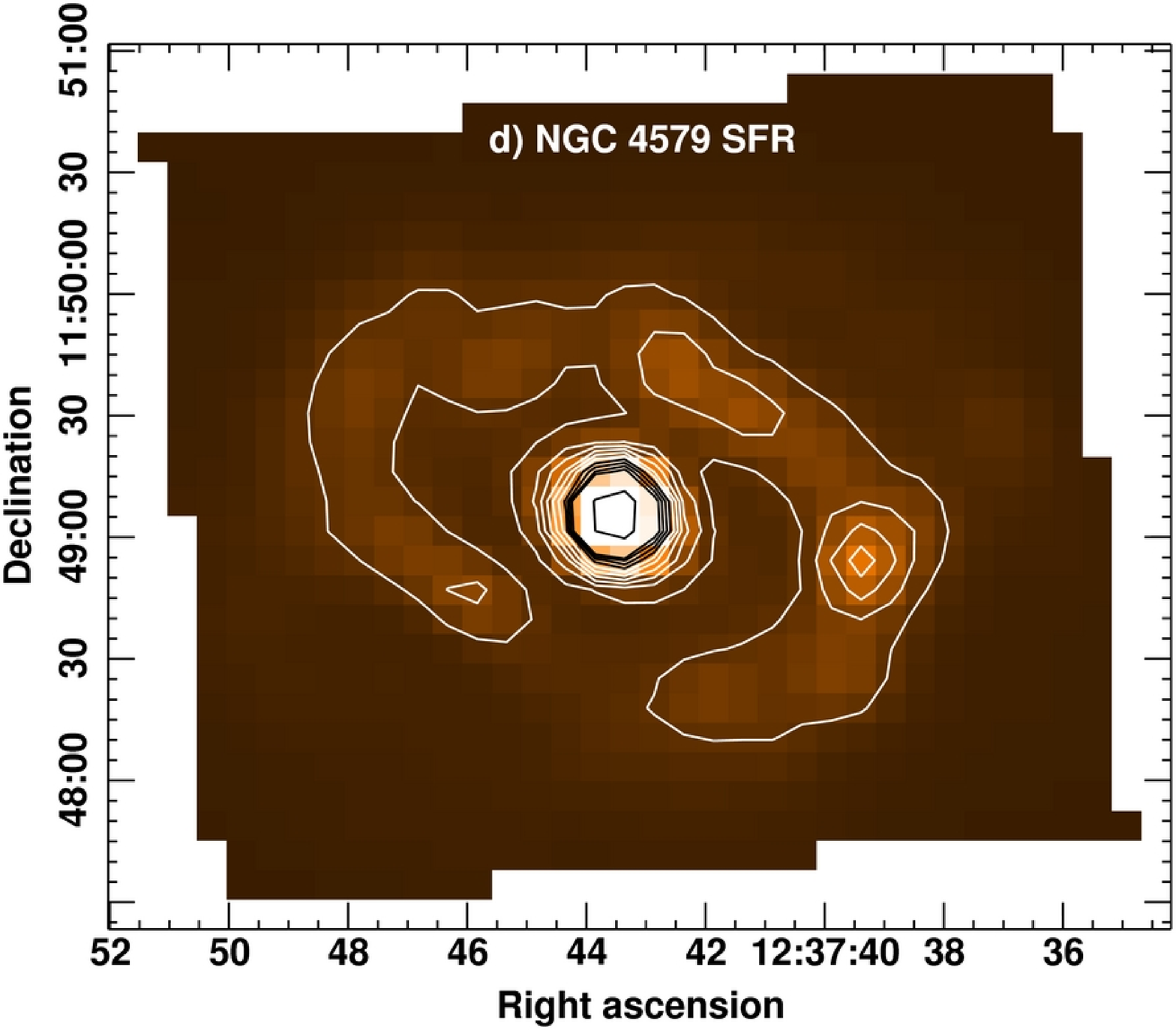}
\caption{(a) Star formation rate surface density for NGC 4254. Contours
  are 8 to 120 by 8 in units of
$10^{-3}$ M$_\odot$ yr$^{-1}$ kpc$^{-2}$ and 
greyscale runs from -6 to $48\times 
  10^{-3}$ M$_\odot$ yr$^{-1}$ kpc$^{-2}$.
 Coordinate epoch is J2000.
(b) Star formation rate surface density for NGC4321. Contours
  are 4 to 40 by 4 and then 40 to 200 by 40 in units of
$10^{-3}$ M$_\odot$ yr$^{-1}$ kpc$^{-2}$ and greyscale runs from -6 to
$48\times 
  10^{-3}$ M$_\odot$ yr$^{-1}$ kpc$^{-2}$.
(c) Star formation rate surface density for NGC4569. Contours
  are 4 to 40 by 4 and then 40 to 200 by 40 in units of 
$10^{-3}$ M$_\odot$ yr$^{-1}$ kpc$^{-2}$ and greyscale runs from -6 to $48\times
  10^{-3}$ M$_\odot$ yr$^{-1}$  kpc$^{-2}$.
(d) Star formation rate surface density for NGC4579. Contours
  are 4 to 40 by 4 and then 40 to 80 by 40 in units of 
$10^{-3}$ M$_\odot$ yr$^{-1}$ kpc$^{-2}$ and greyscale runs from -6 to
$48\times 
  10^{-3}$ M$_\odot$ yr$^{-1}$  kpc$^{-2}$.
Note that the central AGN may dominate the
emission in NGC 4579 and so this image 
does not provide an accurate measure of the star formation rate in the
central region of this galaxy.
\label{fig-sfr}}
\end{figure}

All ancillary data images were aligned with JCMT image orientation and
pixel size using the kappa
 command wcsalign. We calculated an image of
the star formation rate surface density 
for each galaxy by combining the H$\alpha$ and
24 $\mu$m images using the formula in \citet{c07}. 
Maps of the star formation rate for each galaxy are given in
Figure~\ref{fig-sfr}. For NGC 4579, the central AGN dominates the
emission at both H$\alpha$ and 24 $\mu$m. Thus, Figure~\ref{fig-sfr}
does not provide an accurate measure of the star formation rate in the
central regions of NGC 4579.

\section{CO $J$=3-2/$J$=1-0 line ratios and molecular gas
  mass}\label{sec-lineratios} 

\subsection{CO $J$=3-2/$J$=1-0 line ratios}

All calibrations to date of
the conversion from CO luminosity to molecular gas mass have been for
the ground state $J$=1-0 transition \citep{s88}. Thus, the CO
$J$=3-2/$J$=1-0 line 
ratio is an important quantity to measure in 
determining molecular gas masses. In addition, 
the CO $J$=3-2/$J$=2-1 line ratio has been suggested previously to be a
good indicator of temperature in giant molecular clouds, with larger
ratios corresponding to larger temperatures \citep{w97}. 
If the CO $J$=2-1/1-0 line ratio is relatively uniform \citep{l08}, then the CO
$J$=3-2/$J$=1-0 line ratio would also be a good temperature indicator.
Alternatively, the line ratio can be
increased if the average density of the gas is increased 
\citep[see e.g.,][for a discussion of this effect in M82]{pw00}. 

\begin{figure}
\includegraphics[angle=0,scale=.2]{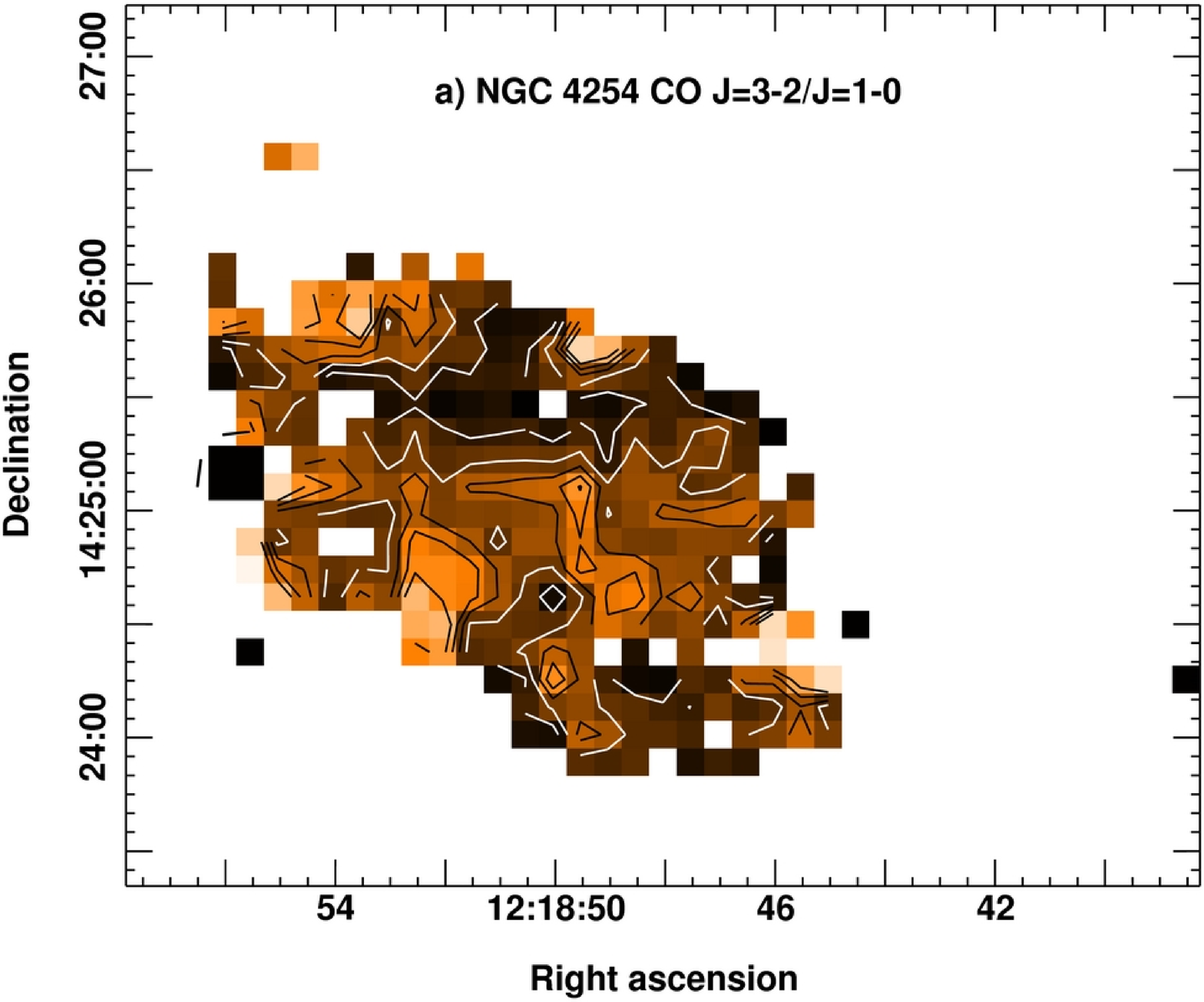}
\includegraphics[angle=0,scale=.2]{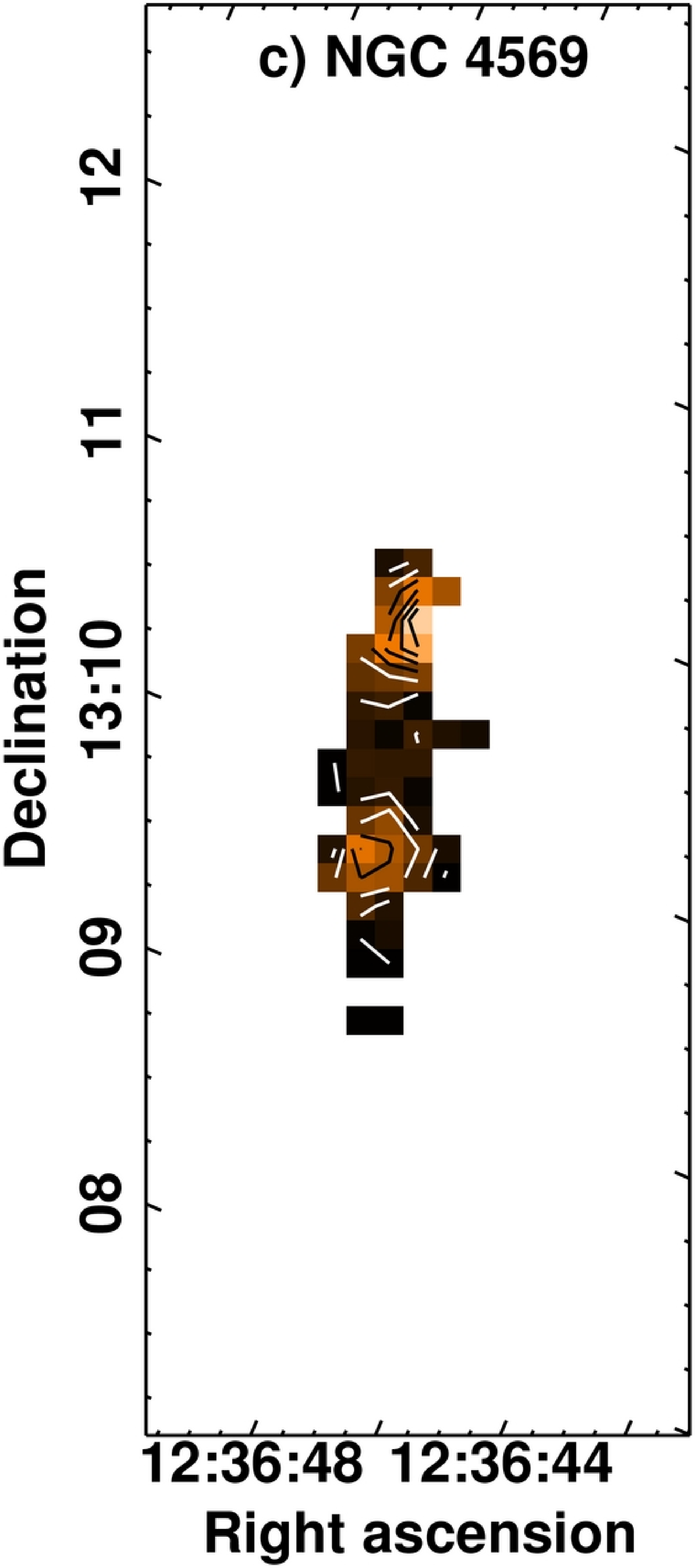}
\includegraphics[angle=0,scale=.2]{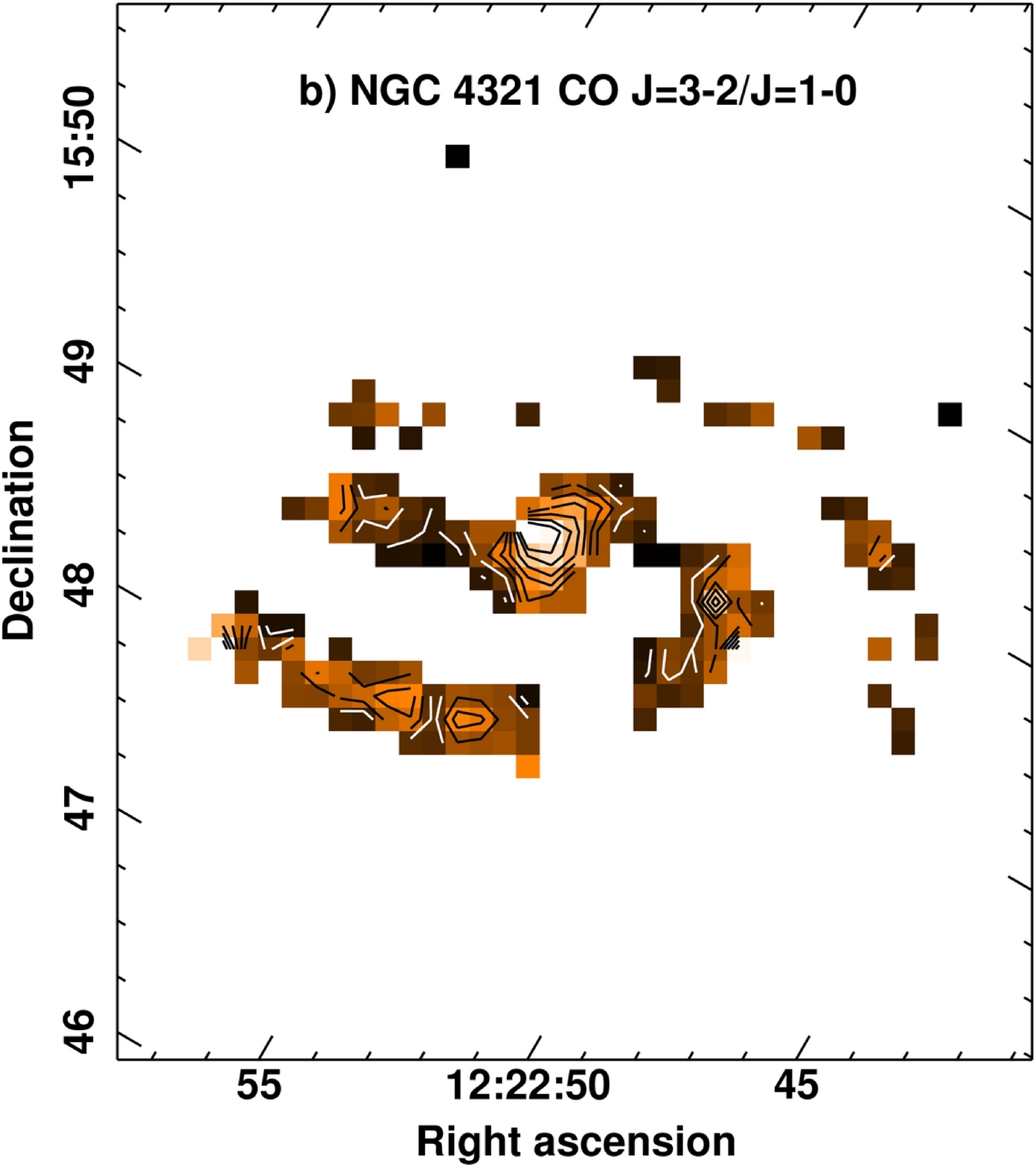}
\caption{(a) CO $J$=3-2/$J$=1-0 ratio for NGC4254. Contours
  are 0.2,0.3 (white), 0.4 to 0.7 by 0.1 (black) and greyscale runs
  from 0.1 to 1 with 
  lighter colors corresponding to higher values.  
 Coordinate epoch is J2000.
(b) CO $J$=3-2/$J$=1-0 ratio for NGC4321. Contours
  are 0.2,0.3 (white), 0.4 to 0.9 by 0.1 (black) and greyscale runs
  from 0.1 to 1.
Note that this map is
oriented at an angle of 30 degrees.
(c) CO $J$=3-2/$J$=1-0 ratio for NGC4569. Contours
  are 0.1,0.2,0.3 (white), 0.4 to 0.7 by 0.1 (black) and greyscale
  runs from 0.1 to 1. 
Note that this map is
oriented at an angle of 23 degrees.
\label{fig-coratio}}
\end{figure}

High resolution CO $J$=1-0 images for NGC 4321 and NGC 4569 have been
published by \citet{h03} and \citet{k07}. 
The total CO $J$=1-0 fluxes measured
directly from these publicly available integrated intensity maps agree
to better than 5\%. However, a comparison of the two images pixel by
pixel shows systematic differences between the two data sets. In
particular, the images from \citet{h03}, which are produced by
combining interferometric and single dish data, 
yield higher
intensities around bright and more compact structures compared to the
images from \citet{k07}. As such systematic differences could have
been produced in the combination of the single dish and
interferometric data (which is a tricky and non-linear process), we
have chosen to calculate the CO $J$=3-2/$J$=1-0 line ratios using the data
from \citet{k07}.
Images of the line
ratios are shown in Figure~\ref{fig-coratio}. In calculating the
average values for each galaxy and for regions within each galaxy, we
limited our analysis to pixels with a signal-to-noise in the CO $J$=3-2
line greater than 3. Global and central values for the line ratios as well as
line ratios for particular emission peaks within the disk are given in
Table~\ref{tbl-lineratios}. 

NGC 4254 has a remarkably uniform CO $J$=3-2/1-0 ratio of 0.33.
NGC 4321 has a similar disk-averaged CO $J$=3-2/1-0 ratio of 0.36;
however, the average 
value in the central 9 pixels (22$^{\prime\prime}$ diameter) is
significantly larger at 0.79.
NGC 4569 shows a significant gradient
from north to south in its line ratio, from 0.53 at the northern CO
$J$=3-2 peak to just 0.06 at the southern peak. Its disk averaged
value of 0.25 is somewhat lower than the other two galaxies.
Finally, NGC 4579 is a difficult case. It is detected weakly by \citet{k07} but
is only marginally
detected at CO $J$=3-2 (Fig.~\ref{fig-co32}). 
Our marginal detection near the
star formation peak in the disk is $0.9 \pm 0.3$ K km s$^{-1}$, while
\citet{k07} detect CO $J$=1-0 emission of 6 K km s$^{-1}$. 
Combining these two values 
suggests that the CO $J$=3-2/$J$=1-0 line ratio in the disk of NGC
4579 is 
similar to the low line ratio seen in the southern portion of NGC
4569.

Assuming a 15\% absolute calibration uncertainty for both CO
data sets (\S2), the mean line ratios in the three 
galaxies with good CO $J$=3-2 detections agree at the 1 sigma level.
However, the southern half of NGC 4569 has a significantly
lower line ratio while the central region of NGC 4321 and the
northern emission peak of NGC 4569 have significantly higher line ratios.
These line ratios 
suggest that the gas may be on average somewhat warmer and/or denser
in the center of NGC 4321 and north of NGC 4569, while the southern
disk of NGC 4569 may be somewhat 
cooler or less dense than typical disk regions. An intriguing
possibility is that the gradient in gas properties in NGC 4569 is
related to its interaction with the ICM; we discuss this possibility
in \S\ref{sec-discuss} below.

The average CO $J$=3-2/1-0 line ratios in these three galaxies are
somewhat smaller than 
the range of values (0.4-0.8) seen in individual giant molecular clouds in M33
 \citep{tw94,w97}.
The line ratios do fall
within the range of ratios (0.2-0.7) measured with similar angular resolution
by  \citet{m99}, albeit at the low end of the
range. However, \citet{m99} observed only the central
21$^{\prime\prime}$ of each galaxy in their sample. If many galaxies
have enhanced line ratios in the central region, as we see in NGC 4321
and possibly in NGC 4569,
this focus on the central regions
could explain their somewhat higher line ratios. Our line ratio 
for NGC 4569 is in 
reasonable agreement with the lower-resolution results
from \citet{hs03}, who measured a line ratio
of $0.23\pm 0.04$. However, our line ratios for the other two galaxies
are substantially larger than the values measured by them: 
$0.21 \pm 0.04$ for NGC 4254 and $0.24\pm 0.02$ for NGC 4321.
Applying a 3$\sigma$ clip to the CO $J$=3-2 data has had the effect of
limiting our line ratio analysis in these galaxies to brighter
regions; for example, the inter-arm regions of NGC 4321 are not
detected at this level. If the interarm gas has a substantially lower
line ratio, this could explain the difference between our values and
those obtained by \citet{hs03}.

\subsection{Molecular gas mass}

The CO $J$=3-2 luminosities and molecular hydrogen gas masses are given
in Table~\ref{tbl-mass}. Values for the center of NGC 4579 are
calculated using a 
3$\sigma$ upper limit and a gaussian line with full-width
half-maximum of 200 km s$^{-1}$. We give two different estimates for the
molecular gas mass. The first value is calculated directly from the CO
$J$=3-2 luminosity 
assuming a CO $J$=3-2/1-0 line ratio of 0.6
and adopting a CO to H$_2$ conversion factor of
$2\times 10^{20}$ cm$^{-2}$ (K km s$^{-1}$)$^{-1}$ \citep{s88}.  
The CO $J$=3-2/1-0 line ratio of 0.6 is adopted as a typical ratio
appropriate to the 
gas in giant molecular clouds. It is
based on observations of an average CO $J$=2-1/1-0 ratio of 0.8 in
spiral disks \citep{l08} combined with the average CO $J$=3-2/$J$=2-1
ratio of 0.76$\pm$0.19 measured in the Galactic molecular cloud M17
\citep{w99}. A similar line ratio ($0.60 \pm 0.13$) is observed in the
central regions of 15 nearby galaxies with modest starbursts
\citep{is08,is09}.
Adopting a line ratio specifically appropriate for
molecular clouds essentially uses the CO $J$=3-2 emission to focus in on
the star forming molecular gas; the CO $J$=1-0 line has been shown to include
emission from lower opacity, more diffuse molecular gas
\citep{ww94,rb05} which likely does not participate directly in massive
star formation.

Alternatively, we could adopt the average CO $J$=3-2/$J$=1-0 line ratio
measured in our sample, which is 0.34 averaged over the three galaxies
in our sample. 
This mass estimate essentially measures the total
molecular hydrogen content and should be directly comparable (modulo
issues of area coverage and sensitivity) to previous estimates of the
molecular gas mass derived directly from CO $J$=1-0 maps.
Table~\ref{tbl-mass} also gives masses measured directly from the data of
\citet{k07} as well as the masses estimated by \citet{h03} from their
single-dish maps. For NGC 4254 and NGC 4321, the CO $J$=1-0 masses
from \citet{k07} are in
general agreement with the masses calculated from the CO $J$=3-2 line
when the line ratio of 0.34 is adopted, as expected given the quite
uniform line ratio seen in these galaxies. However, for NGC 4569, the
CO $J$=1-0 mass is almost a factor of three times larger than the CO
$J$=3-2 mass. This larger mass is partly a result of the fact that the
CO $J$=3-2/$J$=1-0 ratio in this galaxy is the lowest of the three
(accounting for a factor of 1.4 in mass) and also that the southern
region of the galaxy is quite bright in CO $J$=1-0 but almost invisible
in CO $J$=3-2.
For NGC 4579, the CO $J$=1-0 mass is a factor of ten larger than the 
upper limit to the CO
$J$=3-2 mass, again consistent with the very weak CO $J$=3-2 emission in
this galaxy as a whole.

How are we to understand the variety of gas masses given in
Table~\ref{tbl-mass}? The most straightforward interpretation is that
the gas mass calculated from the CO $J$=3-2 luminosity using a CO
$J$=3-2/$J$=1-0 line ratio of 0.6 (as seen in Galactic GMCs) traces the
dense molecular gas, while the mass calculated directly from the CO
$J$=1-0 luminosity traces the total molecular gas mass. With this
interpretation, the dense gas fraction in the galaxies in our sample
is 46\% for NGC 4254, 65\% for NGC 4321, 22\% for NGC 4569, and
$<$7\% for NGC 4579. However, this simple picture is muddied by the
disagreement between the CO $J$=1-0 data of \citet{h03} and
\citet{k07}. Although the total fluxes in the published CO $J$=1-0
integrated intensity images for NGC 4321 and NGC 4569 are very
similar, the {\it total} mass from the single-dish maps of \citet{h03}
is a factor of 2 larger in NGC 4321 and a factor of 0.8 smaller in NGC
4569, such that the dense gas fraction in the two galaxies becomes
quite similar if we use the masses from \citet{h03}. While some of
these variations may be attributed to 
such factors as calibration uncertainties (to explain NGC 4569) and
larger mapping area (for NGC 4321), the fact remains that considerable
uncertainties, perhaps as large as a factor of two, continue to plague
these gas mass estimates.

\section{Gas depletion times}\label{sec-tgas}

The instantaneous gas depletion time ($t_{gas}$) can be calculated as
the ratio of 
the molecular gas mass divided by the star formation rate. The inverse
of the gas depletion time is often referred to as the star formation
efficiency \citep{ys91}. In this section, we examine the gas depletion times
both within and between the galaxies in our sample. 
We focus first on the gas depletion
time calculated assuming a CO $J$=3-2/$J$=1-0 ratio of 0.6, which is
essentially a depletion time for the dense gas. In the next section,
we compare these gas depletion times to values obtained using the CO
$J$=1-0 line.
Note that the {\it instantaneous} gas depletion time (or star
formation efficiency) traces
the rate at which currently available molecular gas is being turned
into stars, and will give (often substantially) shorter times than
global gas depletion times calculated from the total gas reservoir
(molecular plus atomic gas). In the calculations of gas depletion
time, we include a factor of 1.36 to take into account the mass
contribution of helium.

\begin{figure}
\includegraphics[angle=0,scale=.15]{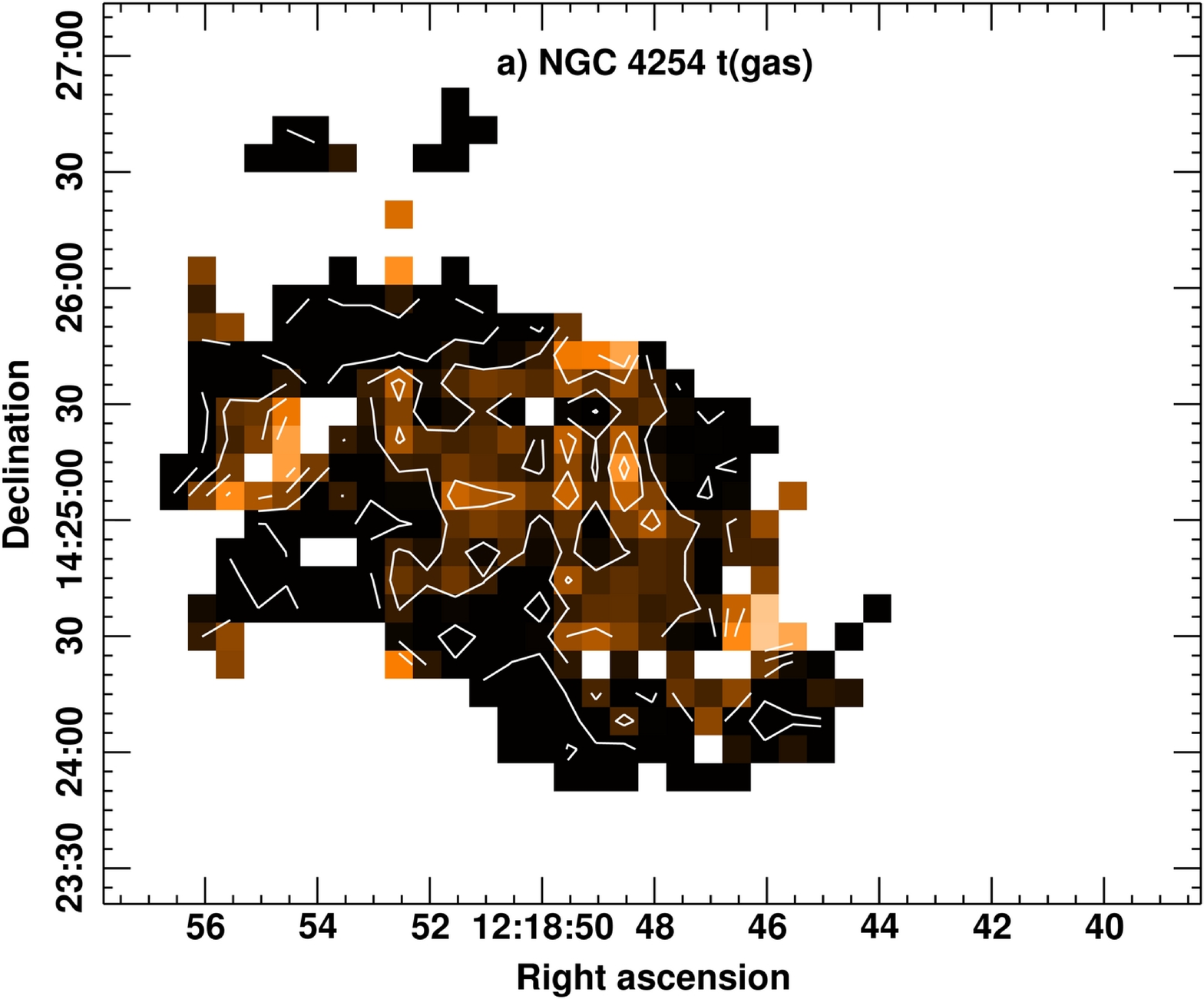}
\includegraphics[angle=0,scale=.1]{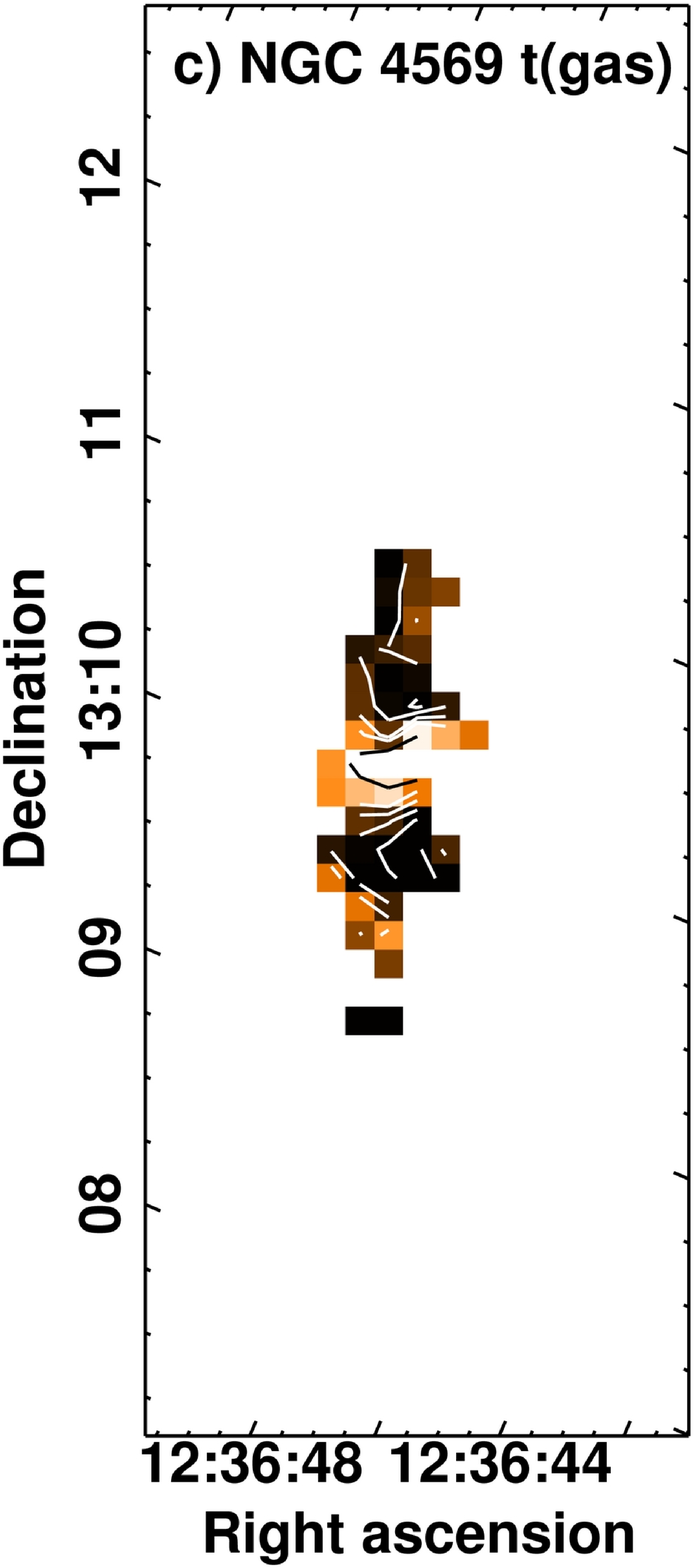}
\includegraphics[angle=0,scale=.15]{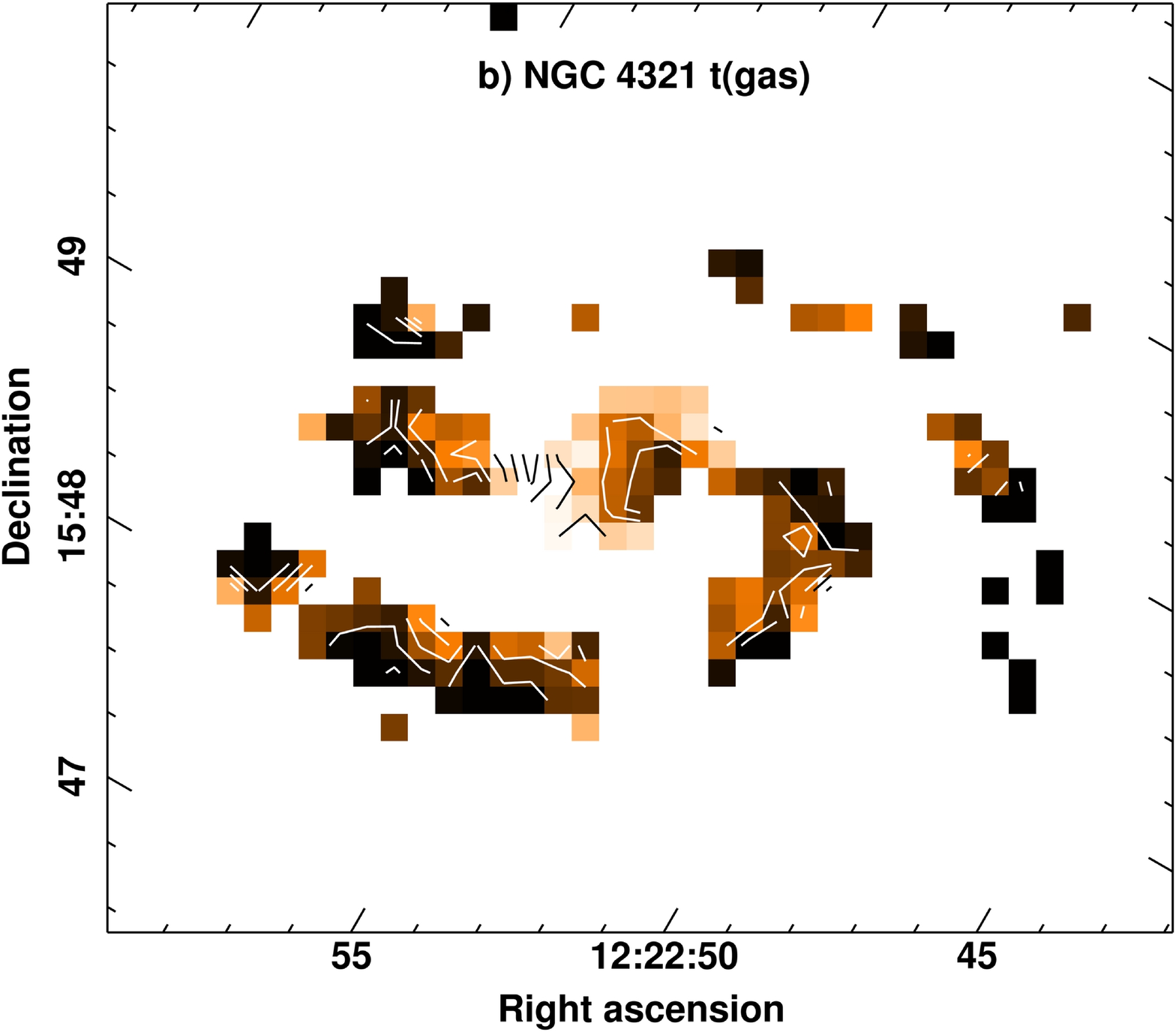}
\caption{
(a) Gas depletion time calculated from the CO $J$=3-2 intensity
  for NGC4254. Contours 
  are 0.8 to 2.0 Gyr by 0.4 Gyr  and greyscale runs from 1 to 3 Gyr
 with
  lighter colors corresponding to higher values.  
(b) Gas depletion time calculated from the CO $J$=3-2 intensity
  for NGC4321. Contours 
  are 0.8 to 2.0 Gyr by 0.4 Gyr 
(white), 3, 4.5, 6 Gyr (black)  and greyscale runs from 1 to 3 Gyr
Note that this map is oriented at an angle of 30 degrees.
(c) Gas depletion time calculated from the CO $J$=3-2 intensity for
NGC4569. Contours 
  are 0.8 to 2.0 Gyr by 0.4 Gyr 
(white)  and 3 Gyr (black)  and greyscale runs from 1 to 3 Gyr
Note that this map is oriented at an angle of 23 degrees.
\label{fig-tgas}}
\end{figure}

We made maps of $t_{gas}$ for NGC4254, NGC 4321, and NGC 4569 by combining
the star formation rate maps described in \S 2 with CO $J$=3-2
integrated intensity maps containing only pixels detected at the 
3$\sigma$ level or above. We assume a CO line ratio of 0.6
and adopt a CO to H$_2$ conversion factor of
$2\times 10^{20}$ cm$^{-2}$ (K km s$^{-1}$)$^{-1}$ \citep{s88}. Maps
of the gas depletion time are 
given in Figure~\ref{fig-tgas}, where the units of $t_{gas}$ are Gyr.
Global and central values for the gas depletion times as well as
values for emission peaks within the disk are given in
Table~\ref{tbl-tgas}. 

For NGC 4254, the average value of $t_{gas}$ is 1.1
Gyr.
For NGC 4321, the average value of $t_{gas}$ is 1.7
Gyr,
while for NGC 4569 it is 1.6 Gyr,
both values 50\% larger than the gas depletion time
for NGC 4254. The gas depletion times show the most variation in NGC
4569, but are 
still quite uniform across the
disk. The
value in the central region  (0.9 Gyr) is
about half the average value in the disk.
The value measured at the peak of
the CO emission in the southern half of
the galaxy is somewhat lower still at 0.6 Gyr, 
while the value at the northern CO peak is closer to the disk mean
value  at 1.2 Gyr.
Note that the weak CO $J$=3-2 emission in the
southern half of the galaxy is balanced by the lower star formation
rate to give a similar gas depletion time.
For NGC 4579, the nominal gas depletion time 
for the one pixel to the west of the nucleus
where the weak CO emission aligns with a region of star formation is
0.25 Gyr, a factor of four smaller than 
the average gas depletion time in NGC 4254 and a factor of six smaller than
the gas depletion times for NGC 4321 and NGC4569. Because of the
strong AGN in NGC 4579 (see \S\ref{sec-discuss}), we cannot obtain a
reliable estimate of 
the gas depletion time in the nucleus from our analysis.

Thus, of the three galaxies in our sample with strong CO $J$=3-2
emission, NGC 4254, which has a normal 
HI content for 
its type, has a smaller gas depletion time by a factor of 1.5 compared to
NGC 4321 and NGC 4569. These two galaxies have
very similar instantaneous depletion times for the dense gas, despite
having a factor of three difference in their HI deficiency
(Table~\ref{tbl-props}). The gas depletion times in these three
galaxies comparable to or somewhat
smaller than the values of $\sim 2\times 10^9$ yr obtained in spiral
galaxies by \citet{l08} and \citet{b08}.
Overall, the gas depletion times calculated here are significantly
smaller than a Hubble time. One reason is that we are calculating an
instantaneous gas depletion time by focusing just on the dense
molecular component rather than including the entire gas reservoir
(atomic plus all phases of molecular gas at all radii). Radial infall
of gas from the outer atomic disks can provide new fuel for star
formation; however, this process is likely to be less effective in
cluster galaxies where the outer disk has been removed \citep{l80}.
Another reason
that these numbers should not be taken as evidence that these Virgo
spirals will imminently ``run out'' of gas to fuel star formation is
that we have not attempted to include the effect of return of material
to the ISM via supernovae, stellar winds, and stellar evolution. 
These
gas depletion times provide a useful snapshot of the current rate at
which gas is being turned into stars, in the same way that the
H$\alpha$ and 24 $\mu$m emission captures the formation of massive ($>
10$ M$_\odot$) stars over the last 10-20 Myr.

\section{The broader implications for molecular gas and star formation}\label{sec-discuss}

\subsection{The effects of internal structure and external environment}

How do the variations in the gas and star formation properties of
these four galaxies relate to their internal structure and larger
environment? NGC 4254 has 
been suggested to be on its first encounter with the central regions
of the cluster and, indeed, has a normal ratio of molecular to atomic
gas and a normal extent of star formation. 
It is also the only galaxy in our sample that does not show evidence
of a bar \citep{l02}.
Its smaller gas depletion
time translates into a higher current star formation efficiency; this
higher efficiency may be due to the fact that the galaxy has only
begun to be disturbed by the cluster \citep[e.g.,][]{h07}. 

NGC 4321 is classified as weakly
barred 
\citep[SAB(s)bc,][]{buta07}; high resolution CO $J$=1-0 observations
show two spiral 
arms and gas in the nucleus as well as faint emission bridging the
arms and the nucleus \citep{gb05}. There is also a prominent
circumnuclear ring 
with enhanced star formation \citep{k95a,k95b}. Thus, the similar gas
depletion times calculated for the central region and the spiral arms
suggests that increased star formation is balanced by
an increased molecular gas content in these regions. 

For NGC 4569, the somewhat lower CO $J$=3-2/$J$=1-0 ratio
of 0.25 suggests that the average gas in NGC 4569 may be somewhat cooler and/or
less dense than the gas in NGC 4254 or 4321. However, this potential
difference in 
average gas properties does not seem to have a strong effect on the
gas depletion time 
(or star formation efficiency) estimated using the CO $J$=3-2 line. Given
the nearly edge-on orientation of NGC 4569, it is possible that the
contribution from CO $J$=1-0 emission in the outer disk of the galaxy
contributes significantly to the measured line ratio, while the CO
$J$=3-2 is confined to the inner disk regions along with the star
formation. 
NGC 4569 is the largest of the Virgo spirals in our sample,
yet its CO $J$=3-2 emission is the most compact. This morphology suggests that
whatever processes are acting to 
strip the outer portions of the galaxy of gas are also affecting the
distribution of the dense molecular gas, perhaps by stripping it
directly or by removing the reservoir (atomic hydrogen or diffuse
molecular hydrogen) from which it forms. 
The southern portion of the
galaxy contains much less dense molecular gas and star formation than
the northern half. Given the geometry of the galaxy \citep{v04}, the
northern portion seems likely to be more affected by interaction with
the IGM as its rotational motion and motion through the cluster are in
the same sense.  The higher density or warmer gas in the northern half
of NGC 4569 could possibly be due to gas compression in this region
where ram pressure is active.



NGC 4579 appears to be the most unusual galaxy of our small
sample. Its disk has the smallest gas depletion time (highest star formation
efficiency) of any of the four galaxies in our sample. The relatively
weaker star formation in this galaxy as well as its more limited
spatial extent (Figure~\ref{fig-sfr}) suggests a galaxy which may be
exhausting its fuel for star formation. Its relatively early
morphological type 
\citep[SAB(rs)ab,][]{buta07}
is consistent with this picture,
although NGC 4569 has an identical classification.
However, the higher inclination of NGC 4569 means that a given surface
density of molecular gas is more easily detected than in NGC
4579. Indeed, the relatively weak CO J=3-2 emission in the southern
half of NGC 4569 would be undetectable if it were at the same
inclination as NGC 4579. In addition, the broader lines 
(full-width zero intensity 400-500 km s$^{-1}$) seen in the central
region of NGC 4579 by 
\citep{gb08} also imply that its central CO emission is harder to
detect than the central emission in NGC 4569. Thus, NGC 4579 may not
be such an unusual galaxy for its type, but simply observed at a less
favorable inclination angle than NGC 4569.


\subsection{Which CO transition traces star forming gas best?}


In \S\ref{sec-tgas}, we calculated gas depletion times by combining
directly the CO $J$=3-2 intensity with the star formation rate.
If instead we use the CO $J$=1-0 intensity from \citet{k07}, and study
exactly the same regions in each galaxy as in \S\ref{sec-tgas}, the 
picture becomes more complicated. Values for the gas depletion time
calculated using the CO $J$=1-0 line are also 
given in Table~\ref{tbl-tgas}.


For NGC 4254, the gas depletion times remain very uniform when the CO
$J$=1-0 line is used instead of the $J$=3-2 line to trace the molecular
gas. The average gas depletion time shows a larger dispersion, which
hints at more local variations than are seen 
using the CO $J$=3-2 line; however, this could also be due to low
signal-to-noise CO $J$=1-0 spectra being included in our analaysis.


For NGC 4321,  when the CO $J$=1-0 line is
used, the
line ratio is also quite uniform overall but
the gas depletion time for
the central 
region is now roughly twice as
small as that of the disk. The
gas depletion times in the CO 
peaks associated with the spiral arms are intermediate between these
two extremes. The disk-averaged gas depletion time
remains about 50\% larger than that of NGC 4254 whether the CO $J$=1-0
or $J$=3-2 transition is used to trace the molecular gas.

For NGC 4569, the gas
depletion times now show significant variations, ranging from 9.0
Gyr at the southern CO $J$=3-2 peak
to 1.8 Gyr for the central region.
The disk averaged value of 5.4 Gyr, with its
larger relative dispersion, also suggests more variation in the gas
depletion time derived from the CO $J$=1-0 line, and is now almost 3 times
larger than the gas depletion time in NGC 4254.

Thus, for these three galaxies, using the CO $J$=1-0 line to calculate
the gas depletion time instead of the $J$=3-2 line results in a greater
spread in gas depletion times both within galaxies and from one galaxy
to another. If we assume that the star formation efficiency (and hence
gas depletion time) should not vary
significantly from one normal spiral galaxy to another, then 
this analysis suggests that the CO $J$=3-2
line may be a more direct tracer of the dense, star forming molecular gas
than is the CO $J$=1-0 line. 
This result is consistent with the
analysis of \citet{i08}, who find a nearly linear correlation between
the CO $J$=3-2 luminosity and the far-infrared luminosity (as a tracer
of star formation) over 5 orders of magnitude in galaxy luminosity.
The good correlation between the star formation rate surface density
and the CO $J$=3-2 emission suggests that the CO $J$=3-2 transition probes
sufficiently high gas densities to be a good tracer of the dense
cores within molecular clouds that are forming stars with a high
efficiency. In contrast, the CO $J$=1-0 line traces gas with a wider
range of densities, including more diffuse material that is not
directly involved in star formation.  
In this interpretation, using the CO
$J$=3-2 line results in a more accurate
measurement of the instantaneous gas depletion time. 

An alternative interpretation would be that the star forming molecular
gas (and the gas depletion time) is
in fact well-traced by the CO $J$=1-0 line, and the variations seen both
within and between galaxies are real. 
In this case, regions with
higher star formation efficiencies (or 
smaller gas depletion times) would be those with a higher rate of
energy injected due to recent star formation per unit mass of
molecular gas. This higher rate of energy injection could warm the
molecular gas, producing brighter CO $J$=3-2 emission and leading to
a more uniform apparent gas depletion times when the CO $J$=3-2
emission line is used, as observed. 
While such heating processes are
undoubtedly at work in the ISM, we argue that the interpretation of
the CO $J$=3-2 line as a direct tracer of dense star-forming gas
is the more straightforward explanation.

\section{Conclusions}\label{sec-concl}

We have used the JCMT to map the CO $J$=3-2 emission from four spiral
galaxies in the Virgo cluster. These galaxies are included in the
SINGS survey \citep{k03} and so have a wealth of information on their
dust and interstellar medium properties. These galaxies also allow us
to probe a range of galaxy properties including
HI deficiency, star formation rate, and nuclear
activity.

Bright CO $J$=3-2 emission is detected over the extended disks of NGC
4254, NGC4321, 
and NGC 4569; NGC 4579 is weakly detected (3$\sigma$) at one position
in the disk and has only an upper limit in the 
central region.
Combining our data with CO $J$=1-0 maps from \citet{k07}, we derive
disk-averaged CO
$J$=3-2/$J$=1-0 line ratios of 0.33 in NGC 4254, 0.36 in NGC 4321, and
0.25 in NGC 4569. The line ratio in the center of NGC 4321 is
significantly larger (0.79), while NGC 4569 shows a gradient in the
line ratio from 0.06 in the south to 0.53 in the north. 
The very weak
CO $J$=3-2 emission in NGC 4579 suggests a line ratio in the disk
similar to the lowest
value seen in NGC 4569. These line
ratios are towards the low end of the range typically measured in
galaxies \citep{m99}. However, previous observations have focused on
the extreme central regions, which our data suggest may have enhanced
line ratios in many galaxies.

We combine H$\alpha$ and 24 $\mu$m images from the SINGS survey
\citep{k03} to obtain maps of the star formation rate following the
prescription in \citet{c07}. We combine these star formation rate
images with maps of the molecular gas surface density derived from the
CO $J$=3-2 data to obtain maps of the instantaneous gas depletion
timescale (the inverse of the star formation efficiency).
Of the three CO-bright galaxies, 
the disk-averaged gas depletion time is shortest for NGC 4254
(1.1 Gyr), roughly 50\% less than for NGC 4321 (1.7
Gyr) and NGC 4569 (1.6 Gyr). 
The fourth galaxy, NGC 4579, has 
weak CO $J$=3-2 emission and small gas
depletion times in the disk, 4-6 times smaller than in the other three
galaxies. 
These times, which are substantially
shorter than a Hubble time, are a measure of the instantaneous gas
depletion time and do not take into account possible replenishing of
the dense molecular gas from more diffuse molecular or atomic gas.

We have compared the gas depletion times obtained using the CO $J$=3-2
emission with those obtained using the CO $J$=1-0 emission, which is the
more commonly used tracer of molecular gas. We find that
using the CO $J$=3-2 line results in more uniform
instantaneous gas depletion times both within galaxies and from one
galaxy to another. We argue that the CO $J$=3-2 line is a better tracer
of the dense molecular gas involved in star
formation than the CO $J$=1-0 line. Alternatively, real variations in
the star formation efficiency may result in variation in the amount of
heating per unit gas, leading to more uniform apparent gas depletion
times using the CO $J$=3-2 line. These two scenarios may be testable
using radiative transfer codes and multiple CO emission lines to
measure the average physical conditions in the molecular gas.

Looking at the internal structure and external environment,
we suggest that the smaller gas
depletion time (or larger star formation efficiency) in NGC 4254 may
be due to the fact that it seems to be encountering the dense portion of the
Virgo cluster for the first time \citep{h07}. 
In NGC 4569, the gradient in the CO $J$=3-2/$J$=1-0 line ratio (and more
weakly in the gas depletion time) from south to north suggests that
active ram pressure in the northern half of the galaxy may be
producing an increase in the average density or temperature of the
molecular gas.


\acknowledgments

We thank the anonymous referee for a very useful referee report.
The James Clerk Maxwell Telescope is operated by The Joint Astronomy
Centre on behalf of the Science and Technology Facilities Council of
the United Kingdom, the Netherlands Organisation for Scientific
Research, and the National Research Council of Canada. The research of
J.I., S.C., K.S., and C.D.W. is supported by grants from NSERC
(Canada). A.U. has been supported through a Post Doctoral Research
Assistantship from 
the UK Science \& Technology Facilities Council.
This research has made use of the GOLD Mine Database.
We acknowledge the usage of the HyperLeda database
(http://leda.univ-lyon1.fr).

{\it Facilities:} \facility{JCMT}.




\clearpage


\clearpage

\clearpage

\clearpage

\clearpage

\clearpage

\clearpage






\clearpage

\begin{deluxetable}{lccccccccccccc}
\tabletypesize{\scriptsize}
\tablecaption{Galaxy Properties \label{tbl-props}}
\tablewidth{0pt}
\tablehead{
\colhead{Galaxy} & \colhead{RA(2000)} & \colhead{Dec(2000)} &
\colhead{$V_{hel}$} 
& \colhead{Type\tablenotemark{a}}
& \colhead{$D_{25}$\tablenotemark{a}} 
& \colhead{$i$\tablenotemark{a}}
& \colhead{$B_T$\tablenotemark{a}}
& \colhead{$\log F_{H\alpha}$\tablenotemark{b}}
& \colhead{HI mass\tablenotemark{c}}
& \colhead{HI}
\\
\colhead{} & \colhead{(hms)} & \colhead{($^{o~\prime~\prime\prime}$)}
& \colhead{(km s$^{-1}$)} 
& \colhead{} 
& \colhead{($^{\prime}$)} 
& \colhead{($^o$)} 
& \colhead{(mag)} 
& \colhead{(erg s$^{-1}$ cm$^{-2}$)}
& \colhead{($10^9$ M$_\odot$)} 
& \colhead{deficiency\tablenotemark{b}} 
}
\startdata
NGC 4254 & 12:18:49.6 & 14:24:59 & 2412 & Sa(s)c & 5.4 & 29 & 10.44 
& -10.95 & 4.7 & 0.02 \\ 
NGC 4321 & 12:22:54.9 & 15:49:20 & 1599 & SAB(s)bc & 7.6 & 32 & 10.05
& -11.13 & 2.5 & 0.52 \\ 
NGC 4569 & 12:36:49.8 & 13:09:46 & -179 &  SAB(rs)ab & 10.4 & 63 & 10.26 
& -11.83 & 0.82 & 0.99 \\ 
NGC 4579 & 12:37:43.5 & 11:49:05 & 1540 & SAB(rs)ab  & 5.6 & 37 & 10.48 
& -11.54 & 0.74 & 1.00 \\ 
\enddata
\tablenotetext{a}{From \citet{buta07}}
\tablenotetext{b}{From \citet{k01}.}
\tablenotetext{c}{From \citet{ky89}, adjusted to a distance of 16.7
  Mpc.}
\end{deluxetable}



\begin{deluxetable}{lccccccccccccc}
\tabletypesize{\scriptsize}
\tablecaption{Observational Parameters\label{tbl-obs}}
\tablewidth{0pt}
\tablehead{
\colhead{Galaxy} & 
\colhead{$\Delta X$\tablenotemark{a}} &
\colhead{$\Delta Y$\tablenotemark{a}} & \colhead{PA\tablenotemark{a}}
& \colhead{Date\tablenotemark{b}}   
& \colhead{$\overline{T_{sys}}$}
& \colhead{$\overline{\tau}$\tablenotemark{c}} &
\colhead{$t_{int}$\tablenotemark{d}}  
& \colhead{$\Delta T$\tablenotemark{e}} & resolution\\
\colhead{} 
& 
\colhead{($^{\prime\prime}$)} & 
\colhead{($^{\prime\prime}$)} & \colhead{($^o$)} & \colhead{} & \colhead{(K)} &
\colhead{(225 GHz)} & \colhead{(s)} & \colhead{(mK)} &
\colhead{($^{\prime\prime}$)}   
}
\startdata
NGC 4254 & 162 & 138 & 0 
& 0214,0228,0427 & 472 & 0.10-0.16 & 90 & 14 & 14.5 \\ 
NGC 4321 & 228 & 186 & 30 &
0211,0213 & 318 & 0.04-0.12 & 30 & 15 & 14.5 \\ 
NGC 4569 & 312 & 138 & 23 &
0212,0213 & 380 & 0.07-0.12 & 40 & 15 & 14.5 \\ 
NGC 4579 & 168 & 132 & 95 &
0228,0301 & 373 & 0.06-0.12 & 60 & 13 & 14.5 \\ 
\enddata
\tablenotetext{a}{Size and orientation of mapped region.}
\tablenotetext{b}{Month,day; all observations done in 2008.}
\tablenotetext{c}{Range of atmospheric optical depth at 225 GHz on
dates when data were obtained.}
\tablenotetext{d}{Typical integration time per point in the map.}
\tablenotetext{e}{Rms noise in spectra at 20 km s$^{-1}$ resolution on
  $T_A^*$ scale; divide by 0.67 for $T_{MB}$.}
\end{deluxetable}



\begin{deluxetable}{lccccccccccccc}
\tabletypesize{\scriptsize}
\tablecaption{CO $J$=3-2/$J$=1-0 line ratios\label{tbl-lineratios}}
\tablewidth{0pt}
\tablehead{
\colhead{Region} &\colhead{NGC 4254} & 
  \colhead{NGC 4321} &
\colhead{NGC 4569} & 
\colhead{NGC 4579} 
}
\startdata
Global & 0.33$\pm$0.15\tablenotemark{a} & 0.36$\pm$0.19 & 0.25$\pm$0.16 & ... \\
Central\tablenotemark{b} & 0.43$\pm$0.11 & 0.79$\pm$0.21 & 0.34$\pm$0.11 & ... \\
Bar ends & ... & 0.42$\pm$0.16 & 0.53$\pm$0.18\tablenotemark{c} & ... \\ 
  & ... & 0.36$\pm$0.12 & 0.06$\pm$0.03 & ... \\ 
Disk peaks & ... & 0.41$\pm$0.12 & ... & 0.15$\pm$0.05 \\
 & ... & 0.31$\pm$0.10 & ... & ... \\
\enddata
\tablenotetext{a}{Average value $\pm$ Standard deviation
}
\tablenotetext{b}{The line ratio measured over the central 8-9 pixels.
}
\tablenotetext{c}{The northern half of NGC 4569 has the higher line ratio.
}
\end{deluxetable}


\begin{deluxetable}{lccccccccccccc}
\tabletypesize{\scriptsize}
\tablecaption{Molecular gas masses\label{tbl-mass}}
\tablewidth{0pt}
\tablehead{
\colhead{Galaxy} &\colhead{SFR\tablenotemark{a}} & 
  \colhead{$L_{CO}(3-2)$} &
\colhead{$M_{H_2}$\tablenotemark{b}} & 
\colhead{$M_{H2}$\tablenotemark{c}} &
\colhead{$M_{H_2}$\tablenotemark{d}} & 
\colhead{$M_{H_2}$\tablenotemark{e}}  
\\ 
\colhead{} &
\colhead{(M$_\odot$ yr$^{-1}$)} & \colhead{(10$^8$ K km s$^{-1}$ pc$^2$)} 
& 
\colhead{($R$=0.6)} &
\colhead{($R$=0.34)} &
\colhead{(Kuno)} &
\colhead{(Helfer)} \\
\colhead{} &
\colhead{} &
\colhead{} &
\colhead{($10^9$ M$_\odot$)}
& \colhead{($10^9$ M$_\odot$)} & \colhead{($10^9$ M$_\odot$)} 
& \colhead{($10^9$ M$_\odot$)}  
}
\startdata
NGC 4254 & 4.8 & 5.6 & 3.0 & 5.3
& 6.5 & ... \\
NGC 4321 & 2.7 & 3.7 & 2.0 & 3.5 
& 3.1 & 6.50 \\
NGC 4569 & 1.1 & 1.3 & 0.67 & 1.2 
& 3.0 & 2.40 \\
NGC 4579 & $<$0.9 & $<$0.19 
  & $<$0.10 
& $<$0.18  
& 1.4 & ... \\
\enddata
\tablenotetext{a}{Total star formation rate derived from H$\alpha$ and 24
  $\mu$m images shown in Figure~\ref{fig-sfr} and described in
  \S\ref{sec-obs}. Note that the true central star 
  formation rate may be overestimated if
  nuclear activity contributes substantially to the 24 $\mu$m flux;
  this is certainly the case in NGC 4579.
}
\tablenotetext{b}{Mass of molecular hydrogen gas calculated from the
  CO $J$=3-2 luminosity and assuming a CO $J$=3-2/1-0 line ratio of $R$=0.6;
  see \S\ref{sec-lineratios} for details.}
\tablenotetext{c}{Mass of molecular hydrogen gas calculated from the
  CO $J$=3-2 luminosity using the measured average CO $J$=3-2/1-0 line
  ratio of $R$=0.34;
  see \S\ref{sec-lineratios} for details. }
\tablenotetext{d}{Mass of molecular hydrogen gas measured from the CO
  $J$=1-0 images from \citet{k07} and using the 
  CO-to-H$_2$
  conversion factor 
($2\times 10^{20}$ cm$^{-2}$ (K km s$^{-1}$)$^{-1}$)
and distance (16.7 Mpc) adopted in this paper.}
\tablenotetext{e}{Mass of molecular hydrogen gas from \citet{h03}
  measured from the CO $J$=1-0 line using the NRAO 12m telescope,
  adjusted to a distance of 16.7 Mpc.
}
\end{deluxetable}

\begin{deluxetable}{lccccccccccccc}
\tabletypesize{\scriptsize}
\tablecaption{Gas depletion times (Gyr)\label{tbl-tgas}}
\tablewidth{0pt}
\tablehead{
\colhead{Region} &\colhead{NGC 4254} & 
  \colhead{NGC 4321} &
\colhead{NGC 4569} & 
\colhead{NGC 4579} 
}
\startdata
CO $J$=3-2 & \\
\hline
Global & 1.11$\pm$0.02(0.42)\tablenotemark{a} & 1.70$\pm$0.07(0.94) & 1.6$\pm$0.1(0.9) & ... \\
Central\tablenotemark{b} & 1.31$\pm$0.07(0.25) & 2.2$\pm$0.2(0.7) & 0.92$\pm$0.08(0.25) & ... \\
Disk peaks & ... & 1.3-1.6 & 0.60$\pm$0.01(0.01),1.22$\pm$0.07(0.22)\tablenotemark{c} & 0.25$\pm$0.08 \\
\hline
CO $J$=1-0 & \\
\hline
Global & 2.13$\pm$0.08(1.32) & 3.3$\pm$0.2(2.2) & 5.4$\pm$0.6(4.1) & ... \\
Central\tablenotemark{b} & 2.1$\pm$0.2(0.6)& 2.0$\pm$0.3(1.0)   & 1.8$\pm$0.2(0.7) & ... \\
Disk peaks & ... & 2.1-2.5 & 2.0$\pm$0.3(1.0),9.0$\pm$0.6(0.8)\tablenotemark{d} & ... \\
\enddata
\tablenotetext{a}{Average value $\pm$ uncertainty in the mean (standard deviation)
}
\tablenotetext{b}{The line ratio measured over the central 8-12 pixels.
}
\tablenotetext{c}{The northern half of NGC 4569 has the larger gas
  depletion time.
}
\tablenotetext{d}{The southern half of NGC 4569 has the larger gas
  depletion time.
}
\end{deluxetable}

\end{document}